\documentclass[11pt,superscriptaddress]{article}

\usepackage[a4paper, total={5.8in, 9in}]{geometry}
\usepackage[indent,skip=4pt]{parskip}
\usepackage{amsmath,amssymb,amsfonts,amsthm,mathrsfs,mathtools,cases}
\usepackage{dsfont}
\usepackage{authblk}
\usepackage[colorlinks=true,linkcolor=blue, citecolor=blue, bookmarks]{hyperref}
\usepackage{cleveref}
\usepackage{microtype}
\usepackage{comment}
\usepackage{graphicx,adjustbox} 
\usepackage{subfig}
\usepackage{wrapfig}
\usepackage[font=small,labelfont=bf]{caption}
\usepackage[usenames,dvipsnames]{xcolor}  
\usepackage{booktabs}
\usepackage{braket}
\usepackage[nottoc]{tocbibind}
\usepackage{nicematrix}
\usepackage{tikz}
\usepackage{pgfplots}
\usepackage{cite}
\pgfplotsset{compat=1.18}
\usepackage{enumitem}

\usetikzlibrary{matrix}
\usetikzlibrary{decorations.markings}
\usetikzlibrary{patterns}
\usetikzlibrary{arrows.meta}

\numberwithin{equation}{section}

\crefname{figure}{Fig.}{Figs.}
\crefname{equation}{Eq.}{Eqs.}
\crefname{section}{Sec.}{Secs.}
\crefname{appendix}{Appendix}{Appendices}
  
\colorlet{darkerblue}{MidnightBlue!20!black}
\colorlet{lightblue}{blue!70!white}

\hypersetup{
    colorlinks,
    linkcolor={Blue},
    citecolor={Blue},
    urlcolor={Blue},
    backref=true
}

\newcommand{\beq}{\begin{equation}}
\newcommand{\eeq}{\end{equation}}
\newcommand{\beqa}{\begin{eqnarray}}
\newcommand{\eeqa}{\end{eqnarray}}

\newcommand{\Tr}{\operatorname{Tr}}

\usepackage[framemethod=TikZ]{mdframed}

\title{\bf Quench dynamics of negativity Hamiltonians}

\author{Riccardo Travaglino$^1$, Colin Rylands$^{1,2}$ and Pasquale Calabrese$^{1,3}$}

\date{}

\begin{document}
\maketitle
{\small
\vspace{-5mm}  \ \\
{$^{1}$}  SISSA and INFN Sezione di Trieste, via Bonomea 265, 34136 Trieste, Italy\\[-0.1cm]
\medskip{$^{2}$}Centre for Fluid and Complex Systems, Coventry University, Coventry, CV1 2TT, United\\[-0.3cm]
\medskip Kingdom
\\[-0.3cm]
\medskip
{$^{3}$}  International Centre for Theoretical Physics (ICTP), Strada Costiera 11, 34151 Trieste, Italy
}

\begin{abstract}
In this paper, we investigate the quench dynamics of the negativity and fermionic negativity Hamiltonians in free fermionic systems. We do this by generalizing a recently developed quasiparticle picture for the entanglement Hamiltonians to tripartite geometries.   We obtain analytic expressions for these quantities which are then extensively checked against previous results and numerics. 
In particular, we find that the standard negativity Hamiltonian contains both non-local hopping terms and four fermion interactions, whereas the fermionic version is purely quadratic. However, despite their marked difference, we show that the logarithmic negativity  obtained from either are identical in the ballistic scaling limit, as are their symmetry resolution.

\end{abstract}
\newpage

\tableofcontents

\section{Introduction}
The wavefunction of a many-body quantum system contains an overabundance of information, presenting a significant roadblock on the route to understanding  such systems. One of the main tasks of modern quantum statistical physics is to develop quantities and techniques which can distil this wealth of information into a transparent form and provide insight into the physical properties of the system. One such property, unique to quantum systems, is the entanglement between different parts of an extended system~\cite{vidal2003,Plenio2005AnIT,damico2008,horodecki2009,Calabrese_2009_1}. For pure states, this can be elegantly encapsulated in a single number -- the bipartite entanglement entropy between a subsystem $A$ and its complement $\bar{A}$. This is defined to be,
\begin{eqnarray}
    S_A=-\Tr [\rho_A\log (\rho_A)]~,~\rho_A=\Tr_{\bar A}[\rho],
\end{eqnarray}
where $\rho=\ket{\Psi}\!\bra{\Psi}$ is the density matrix of the system in the state $\ket{\Psi}$ and $\rho_A$ the reduced density matrix of the subsystem. By understanding how $S_A$ behaves as parameters of the system or subsystem are changed, one can unveil fundamental and universal properties of many-body systems. 
In equilibrium, one of the most celebrated successes of the entanglement entropy has been the prediction of universally valid laws  for its scaling with subsystem size\cite{arealaws,Pasquale_Calabrese_2004,Calabrese_2009}. Away from equilibrium, the entanglement entropy remains of central importance. This is particularly so, when the system is taken out of equilibrium via a quantum quench. Therein, the build up of entanglement entropy between $A$ and $\bar A$ provides deep insight into the local relaxation towards a stationary state \cite{Deutsch1991,srednicki1,Rigol:2007juv,polkovnikov,DAlessio:2015qtq,Gogolin_2016,Essler_quench,Calabrese_2016,Collura_2014}. As in equilibrium, it also displays universal dynamical behaviour such as a linear increase in time for systems quenched from lowly entangled states~\cite{quench2}. A full characterization of the entanglement between $A$ and $\bar A$ is provided at an operatorial level by the entanglement Hamiltonian, $K_A$, which is related to the reduced density matrix via
\begin{equation}
    \label{eq:ent_Ham_def}\rho_A=\frac{1}{\mathcal {Z}_A}e^{-K_A},~~\mathcal {Z}_A=\Tr[e^{-K_A}].
\end{equation}
As with the full wavefunction, $\rho_A$ contains a huge amount of information which can be packaged in a complicated fashion. Despite this complexity, $K_A$ can display some remarkable and universal properties. One such example is the spatial locality guaranteed by the Bisognano-Wichmann theorem for ground states in quantum field theory~\cite{Bisognano:1975ih,Bisognano:1976za}. This can also be extended to certain non-equilibrium scenarios in the presence of conformal invariance~\cite{Cardy_2016}.  

For mixed states however, $S_A$ is not a good measure of entanglement as it also captures classical correlations between $A$ and $\bar A$. Out of equilibrium, such a scenario arises when $A$ is composed of two distinct parts $A=A_1\cup A_2$ and one wishes to quantify the entanglement between $A_1$ and $A_2$. To more accurately quantify entanglement in such cases, the entanglement negativity was introduced \cite{VidalPRA2002,Plenio2005}.  This is defined as,
\begin{equation}
    \mathcal{E} = \log \Tr |\rho_A^{T_1}|
\end{equation}
where $T_1$ represents the partial transposition inside only $A_1$ and $|\rho_A^{T_1}|=\sqrt{\rho_A^{T_1}\left(\rho_A^{T_1}\right)^\dag}$. This quantity is inspired by Peres criterion \cite{PeresPRL1996,Simon2000}, which states that the partial transpose of a density
matrix can have negative eigenvalues only if the state is entangled.  The negativity effectively counts the amount of negative eigenvalues in the partial transposed density matrix, since 
\begin{equation}
    \Tr |\rho_A^{T_1}| = 1 + 2 \sum_{\lambda_i<0} |\lambda_i|~,
\end{equation}
and therefore provides a signature of entanglement between $A_1$ and $A_2$. 
The entanglement negativity, or negativity for short, has been used to characterize entanglement properties in various setup and models, for instance, in field theories \cite{calabrese_cardy_tonni_2012,Calabrese_cardy_tonni_2013,Capizzi_2022,Capizzi_2022b,Calabrese_2014,Blondeau_2016,Castro_Alvaredo_2019,neg_spec}, in harmonic chains \cite{negativity_harmonic,Cavalcanti_2008,Eisler_2014} 
 random models \cite{negativity_random,spin3,Wybo_2021} and spin chains \cite{spin1,spin2,spin4,Rogerson_2022,Choi_2024}. For free bosonic theories the partial transpose can be related, via the coherent state path integral, to partial time reversal and preserves the Gaussianity of the state. This enables the efficient calculation of the negativity via the Peschel trick for free theories.

Practically, the computation of $\mathcal{E}$  is often carried out by means of a replica trick where one defines the R\'enyi negativites, $\mathcal{E}_\alpha$
\begin{equation}
    \mathcal{E}_\alpha = \log \Tr (\rho_A^{T_1})^\alpha~, ~\alpha\in \mathbb{N}.
\end{equation}
Note that these are not proper entanglement measures, and their usefulness resides mostly in the possibility to re-obtain the usual negativity for $\alpha_e\to 1$ where $\alpha_e$ is even \cite{calabrese_cardy_tonni_2012,Calabrese_cardy_tonni_2013}.
They are experimentally accessible via the randomised measurement toolbox \cite{Elben_2020,Neven_2021,Elben_2022} and some combinations can be used as entanglement witnesses \cite{Elben_2020,Neven_2021}. 
Another interesting object of study are the ratios $R_\alpha$
\begin{equation}
    R_\alpha = \frac{e^{\mathcal{E}_\alpha}}{\Tr [\rho_A^\alpha]}
\end{equation}
which have attracted some interest also because of their universal properties, which allow to use them as signatures of criticality \cite{calabrese_cardy_tonni_2012,Calabrese_cardy_tonni_2013,Calabrese_tagliacozzo_tonni_2013,Alba_2013,chung_alba_2014,Calabrese_2014}.

The negativity has been studied also for fermionic theories 
\cite{shapourian2017many,eisler2016negativity2d,Shapourian2017,Shapourian_2019,shapourian_ruggiero,Cornfeld2019 }. Therein however, it was shown that the partial transpose, implemented in the occupation number basis, does not exhibit certain expected properties of an entanglement measure like additivity and can fail to capture some correlations of Majorana fermions~\cite{Shapourian2017}. Moreover, it was shown that due to fermion statistics, the partial transpose of a Gaussian state results in a sum of two generically non-commuting Gaussian states. To alleviate these issues an alternative quantity, dubbed the fermionic negativity, was introduced~\cite{Shapourian2017,shapourian2017many}. This is given by, 
\begin{equation}
    \mathcal{E}^{(f)} = \log \Tr |\rho_A^{R_1}|,
\end{equation}
where $R_1$ is a partial time reversal acting on $A_1$. This quantity indeed has the appropriate properties for a measure of entanglement~\cite{shapourian2017many} and additionally preserves the Gaussianity of a state. In fact, partial time reversal of a Gaussian state returns one of the Gaussian states which appears after a partial transposition. As with the previous quantity, $\mathcal{E}^{(f)}$ is often computed using a modified replica trick. In this instance, since the partial time reversal does not preserve Hermiticity~\cite{shapourian2018negativity}, the relevant quantity is 
\begin{eqnarray}
\label{eq:renyinegativitiesdefinition}
    \mathcal{E}_\alpha^{(f)}=\begin{cases}
         \Tr \big[(\rho^{R_1}_A [\rho^{R_1}_A]^\dagger )^{\alpha/2}\big]&\alpha~{\rm even}\\
       \Tr \big[(\rho^{R_1}_A [\rho^{R_1}_A]^\dagger )^{\frac{\alpha-1}{2}} \rho^{R_1}_A\big]  & \alpha~{\rm odd}
    \end{cases},
\end{eqnarray}
with the fermionic negativity once again resulting from the limit $\alpha_e\to 1$ for $\alpha_e$ even. By analogy, $R^{(f)}_\alpha$ can also be defined
\begin{eqnarray}
    R^{(f)}_n=\frac{e^{\mathcal{E}^{(f)}_n}}{\Tr[\rho_A^\alpha]}
\end{eqnarray}
and also investigated for universality. 

Inspired by the entanglement Hamiltonian, the negativity Hamiltonian $\mathcal{N}_A$, has also been introduced and provides an insight in the negativity at an operatorial level. Analogously to~\eqref{eq:ent_Ham_def} it is given by,
\begin{eqnarray}
    \rho_A^{T_1}=\frac{1}{\mathcal{Z}_A}e^{-\mathcal{N}_A}
\end{eqnarray}
and similarly expresses the negativity in an operatorial fashion. In the same way one can also study the fermionic version
\begin{eqnarray}
    \rho_A^{R_1}=\frac{1}{\mathcal{Z}_A}e^{-\mathcal{N}^{(f)}_A}~.
\end{eqnarray}
These quantities have been studied in equilibrium settings~\cite{murcianonegativityHamiltonian,Rottoli_2023,Rottoli2023_finiteT}, displaying some intriguing features akin to those of the entanglement Hamiltonian but with some notable differences. In this work, we shall investigate, for the first time, these two types of negativity Hamiltonian in an out of equilibrium setting.
 
Out of equilibrium, the most successful effective theory for describing the quench dynamics of the entanglement entropy is the quasiparticle picture~\cite{quench2}. 
In this picture, the quench acts on the initial state by producing pairs (or multiplets \cite{bertini2018entanglement,bastianello_pair}) of entangled quasiparticles, which propagate ballistically through the system, thus spreading correlations. Despite the simplicity of the picture, this has been applied with success to the study of several different entanglement-related quantities both in free and interacting models \cite{alba1,alba2,Alba_2019,parez2021quasiparticle,Coser_2014,Murciano2022,Parez_2022,groha2018full,horvath2024full,rottoli2024,travaglino2024quasiparticlepictureentanglementHamiltonians},
although in this last case some issues are still to be clarified \cite{QA,klobas2021,PRX}, since the quasiparticle picture results have not found complete agreement with results obtained through the recently introduced framework of spacetime duality \cite{PRX,PhysRevLett.131.140401,bertini2024dynamics}. Recently, this formalism was applied at the operator level to show the entanglement Hamiltonian itself obeys a quasiparticle picture for certain quenches in one-dimensional free fermionic systems~\cite{rottoli2024}. This was then further generalized to higher dimensions for a wide class of initial states~\cite{travaglino2024quasiparticlepictureentanglementHamiltonians}. 

In this paper we apply the same formalism to the task of calculating the negativity Hamiltonian after a quench in one dimensional free fermionic systems. We consider both types of negativity Hamiltonian using the partial transpose or time reversal. We find expressions for these quantities which are valid in the ballistic scaling limit and check our results against both previous analytic calculations and exact numerics. In particular we show that while standard and fermionic versions of the negativity Hamiltonian are distinct, with the former possessing long range four-fermion interactions, and the latter being purely quadratic, both produce the same results for the negativity and its symmetry resolution~\cite{PhysRevA.98.032302,Murciano_2021}. 

The structure of the paper is as following. In section \ref{sec:setup} we will introduce the Hamiltonians and states of interest together with the setting of the hydrodynamic framework which we will apply in the following. In sections \ref{sec:pure} and \ref{sec:fermionic} we will apply the hydrodynamic framework to obtain the standard and fermionic negativity Hamiltonians, highlighting their relations and differences. Finally, in sections \ref{sec:checks} and \ref{sec:numerics} we provide solid analytical and numerical validation of our results, before drawing conclusions in section \ref{sec:conclusions}.

\section{Setup}
\label{sec:setup}
We start by introducing our system, the quenches we consider and briefly reviewing the quasiparticle picture approach to the entanglement Hamiltonian\cite{rottoli2024}. We consider a one-dimensional free fermionic system, described by a Hamiltonian of the form
\begin{equation}\label{eq:Hamiltonian}
    H =  -\frac{1}{2}\sum_{x=1}^Lc^\dag_xc_{x+1}+c^\dag_{x+1}c_x=\sum_k \varepsilon_k c_k^\dagger c_k ~,
\end{equation}
where $c^\dagger_x,c_x$ are fermioninc creation and annihilation operators for site $x$ and $ c^\dagger_k,c_k$ are their Fourier space counterparts.  Although our discussion is general, we will focus for clarity and numerical comparison on the hopping Hamiltonian, in which $\varepsilon_k = -\cos k$ and the group velocity of the excitations is $v_k = \frac{\partial\varepsilon_k}{ \partial_k} = \sin k$. The model is integrable and has a large number of conserved quantities, including the particle number $N=\sum_{x=1}^Lc^\dag_x c_x=\sum_{k}c^\dag_k c_k$.

We are interested in the quench dynamics of this system and accordingly, it is initialized in some state $\ket{\psi}$, which is not an eigenstate of~\eqref{eq:Hamiltonian}, and allowed to undergo unitary evolution using $e^{-i H t}$. We consider the dynamics emerging from two broad classes of initial states. The first of these is the squeezed state, which in general appears in the so-called mass quenches of free theories~\cite{Fioretto_Mussardo}. It is given by
\begin{equation}\label{eq:squeezed}
    \ket{\psi}= \prod_{k>0}\left( \sqrt{1-n(k)} + e^{i\varphi_{k} }\sqrt{n(k)} c^\dagger_{k} c^\dagger_{-k}\right)\ket{0},
\end{equation} 
where $n(k)=\bra{\psi}c^\dag_kc_k\ket{\psi}=1-\bra{\psi}c^\dag_{k-\pi}c_{k-\pi}\ket{\psi}$ is the fermion occupation number, which is an even but otherwise arbitrary function, $e^{i\varphi_k}$ is an unimportant phase and $\ket{0}$ is the fermion vacuum. This state breaks particle number conservation and is composed of pairs of quasiparticles with opposite momenta endowing it with single site translational symmetry. The (fermionic) tilted ferromagnet and antifeeromagnet, whose quench dynamics have been studied recently, is a state of this type~\cite{ares2023lack,ares2023,Fagotti2014}. 

The other type of state we shall consider preserves particle number symmetry. It is given by
\begin{equation}\label{eq:dimer}
    \ket{\psi}=\prod_{k>0} \left(\sqrt{1-n(k)}c^\dagger_{k-\pi}+e^{i\varphi_k}\sqrt{n(k)} c^\dagger_k \right)\ket{0},
\end{equation}
where again $n(k)$ is the fermion occupation function while the phase $e^{i\varphi_k}$ plays no role for us. This state has two-site translational invariance, and, depending on the choice of $n(k)$, encompasses product states such as the dimer and N\'eel states typically studied in quenches of free fermion models. Here we choose to write it in Fourier space for later convenience and refer to it as the symmetric state. 

\subsection{Quasiparticle picture}
We shall study the quench dynamics from these states in the ballistic scaling regime, in which both the subsystem size and time are sent to infinity $|A|,t\to\infty$ with their ratio held fixed. In this regime, it is natural to take a hydrodynamic approach, which consists, principally, of subdividing the system into mesoscopic hydrodynamic cells of size $\Delta$, and defining partially Fourier transformed operators inside such cells,
\begin{equation}
    b^\dag_{x_0,k} = \frac{1}{\sqrt{\Delta}}\sum_{\varkappa=0}^{\Delta-1}  e^{-ik\varkappa}c^\dagger_{x_0+\varkappa} 
    \label{eq:cellfourier}.
\end{equation}
Here $x_0$ denotes the position of the fluid cell and $\varkappa$ labels the site inside that particular cell. Assuming that the correlations of the initial state decay fast enough on the scale of $\Delta$, then we can express the initial density matrix as
\begin{eqnarray}
    \rho(0)=\prod_{x_0}\prod_{k>0}\rho_{x_0,k}~,
\end{eqnarray}
where $\rho_{x_0,k}$ is the local state within the fluid cell. For the squeezed state this is given by~\cite{Bertini_2018}
\begin{eqnarray}\nonumber
\rho_{x_0,k}&=&n(k)\hat{n}_{x_0}(k)\hat{n}_{x_0,-k}+(1-n(k))(1-\hat{n}_{x_0}(k))(1-\hat{n}_{x_0}(-k))\\
&&+\sqrt{n(k)(1-n(k))}(e^{i\varphi_k}b_{x_0,k}^\dagger b_{x_0,-k}^\dagger +e^{-i\varphi_k} b_{x_0,-k}b_{x_0,k})~,
\end{eqnarray}
with $\hat{n}_{x_0}(k)=b^\dag_{x_0,k}b_{x_0,k}$. Alternatively, for the symmetric state it is~\cite{travaglino2024quasiparticlepictureentanglementHamiltonians}
\begin{eqnarray}\nonumber
    \rho_{x_0,k}{(0)} &=&n(k)\hat{n}_{x_0}(k)(1-\hat{n}_{x_0}(k-\pi)) + (1-n(k))(1-\hat{n}_{x_0}(k) )\hat{n}_{x_0}(k-\pi)\hspace{1cm}\\\label{eq:symm_presrve_pure} &&+ \sqrt{n(k)(1-n(k)) }\left(e^{i\varphi_k}b_{x_0,k}^\dagger b_{x_0,k-\pi}+e^{-i\varphi_k}b_{x_0,k-\pi}^\dagger b_{x_0,k}\right).
\end{eqnarray}
This decomposition is exact for states which have a product structure in real space, like the dimer, N\'eel or tilted ferrmomagnetic states, otherwise it is approximate.  

The quasiparticle picture is a semiclassical effective theory of the quasiparticle dynamics in integrable models. As described in the introduction, it posits that the dynamics is captured by assuming that the quasiparticles are transported ballistically through the system, thereby spreading their initially local correlations. In our case, the quasiparticles are the fluid cell modes created by $b^\dag_{x_0,k}$, and their time evolution is approximated via~\cite{rottoli2024}
\begin{equation}
    b^\dag_{x_0,k}{(t)} = e^{-iHt} b^\dag_{x_0,k}e^{iHt} \approx  b^\dag_{x_t(k),k}
\end{equation}
where $x_t(k)=x_0+v_kt$. Thus, the time evolution ballistically transports the quasiparticles from fluid cell $x_0$ to $x_t(k)$ in a time $t$.  This leads to an effective time-dependent density matrix of the form, 
\begin{equation}
\rho(t)=\prod_{k>0}\prod_{x_0}\rho_{x_0,k}(t)~,
    \label{eq_decomposition}
\end{equation}
where for the squeezed state 
 \begin{multline}
    \rho_{x_0,k}(t) =
    n(k) \hat{n}_{x_t}(k)(k)\hat{n}_{x_t}(-k)    + (1-n(k))(1- \hat{n}_{x_t}(k))(1-\hat{n}_{x_t}(-k) )\\+  \sqrt{n(k)(1-n(k))}(e^{i\varphi_k}b_{x_t(k),k}^\dagger b_{x_t(-k),-k}^\dagger +e^{-i\varphi_k} b_{x_t(-k),-k}b_{x_t(k),k}),
    \label{eq:densitymatrix}
\end{multline}
with $\hat{n}_{x_t}(k)=b^\dag_{x_t(k)}b_{x_{t}(k)}$. Similarly, for the symmetric state we have  
\begin{multline}
    \rho_{x_0,k}{(t)} =n(k)\hat{n}_{x_t}(k)(1-\hat{n}_{x_t}(k-\pi)) + (1-n(k))(1-\hat{n}_{x_t}(k))\hat{n}_{x_t}(k-\pi) \hspace{1cm}\\\label{eq:symm_presrve_pure} + \sqrt{n(k)(1-n(k)) }\left(e^{i\varphi_k}b_{x_t(k),k}^\dagger b_{x_t(k-\pi),k-\pi}+e^{-i\varphi_k}b_{x_t(k-\pi),k-\pi}^\dagger b_{x_t(k),k}\right)\,. 
\end{multline}
We see, therefore, that the state at time $t$  consists of correlated pairs of quasiparticles of opposite momenta which originated from the same fluid cell labelled by $x_0$ but now reside in different ones labelled by $x_{t}(\pm k)$ (or $x_{t}(k), x_{t}(k-\pi)$ for symmetric states). This transport of correlated quasiparticles to different fluid cells is responsible for the spreading of spatial correlations in the system and the growth of entanglement entropy. 
 
 Considering this effective decomposition, the tracing procedure necessary to obtain the reduced density matrix of a subsystem $A$ is greatly simplified. It amounts simply to finding the point to which the operators have been ballistically transported at the time of interest. 
 For example, let us a consider a single pair originating from the $x_0$ cell  and suppose that at time $t$, $x_t(k)\in A$ while $x_t(-k)\notin A$. Then we have that 
 \begin{eqnarray}\label{eq:ent_Ham_modes}
     \Tr_{\overline{A}} \left[ \rho_{x_0,k}{(t)} \right] &=&n(k) \hat{n}_{x_t}(k) + (1-n(k)) (1-\hat{n}_{x_t}(k)) \\
     &=&\frac{1}{1+e^{-\eta(k)}}\exp\left(-\eta(k) \hat{n}_{x_t}(k)\right),
     \label{eq:tracedrho}
 \end{eqnarray}
 where $\eta(k) = \log \frac{1-n(k)}{n(k)}$. This  simple observation allows, with relative ease, to find the entanglement Hamiltonian in bipartite systems. Repeating the procedure for all pairs and fluid cells we are lead to a factorization of the reduced density matrix as 
 \begin{eqnarray}\label{eq:factorized}
     \rho_A(t)&=&\rho_{\rm mixed}(t)\otimes \rho_{\rm{pure}}(t)
 \end{eqnarray}
 with 
 \begin{eqnarray}\label{eq:ent_Ham}
     \rho_{\rm mixed} (t)&=&\frac{1}{\mathcal{Z}_A}e^{-K_A(t)},\\\label{eq:ent_Ham_2}
     K_A(t)&=&\sum_{k}\sum_{\{x_0|x_{t}(k)\in A\,\& \, x_{t}(-k)\notin A\}}\eta(k)\hat{n}_{x_t}(k).
 \end{eqnarray}
Here, the $\rho_{\rm{pure}}(t)$ part comes from quasiparticle pairs which are both inside $A$ while $\rho_{\rm mixed}(t)$ comes from pairs which are shared between $A$ and $\bar A$. The mixed part then gives the entanglement Hamiltonian $K_A$. From this one can straightforwardly Fourier transform $K_A(t)$ back to real space to study its locality and causal structure. Note that the same form applies regardless of the structure of $A$ i.e. connected or disconnected but with the set $\{x_0|x_{t}(k)\in A\,\& \, x_{t}(-k)\notin A\}$ chosen   appropriately.  The pure part, on the other hand, does not contribute to the entanglement between $A$ and $\bar A$ but does contribute to other quantities like the full counting statistics. It cannot be expressed in exponential form, but it can nevertheless be written down explicitly~\cite{travaglino2024quasiparticlepictureentanglementHamiltonians}. 

Written in this way, previous results on the entanglement entropy and other quantities can be re-derived. For example, using the product structure the entanglement entropy is 
\begin{eqnarray}\nonumber
    S_A(t)&=&-\Tr_A [\rho_{\rm mixed}(t)\log \rho_{\rm mixed}(t)]\\
    &=& -\sum_{k}\sum_{\{x_0|x_{t}(k)\in A\,\& \, x_{t}(-k)\notin A\}} \Tr_A\left[\Tr_{\overline{A}} \left[ \rho_{x_0,k}{(t)} \right] \log \Tr_{\overline{A}} \left[ \rho_{x_0,k}{(t)} \right] \right]~,
\end{eqnarray}
where the summand can be determined from~\eqref{eq:ent_Ham_modes} to be 
\begin{equation}
    \Tr_A\left[\Tr_{\overline{A}} \left[ \rho_{x_0,k}{(t)} \right] \log \Tr_{\overline{A}} \left[ \rho_{x_0,k}{(t)} \right] \right]=n(k)\log n(k)+(1-n(k))\log(1-n(k)).
\end{equation} Replacing the sums by integrals, and introducing an appropriate counting function in the spatial integral to account for the restrictions of the $x_0$ sum, the well known quasiparticle expression is obtained~\cite{quench2,fagotti2012evolution}:
\begin{equation}
S_A(t)= \int_{-\pi}^{\pi} \min(2v_kt,\ell)[-n(k)\log n(k)-(1-n(k))\log(1-n(k))].
\end{equation}
where $\ell$ is the subsystem size. 
 \begin{figure}
    \centering
    \begin{tikzpicture}[scale=0.75]
   
        \draw[black,thick,->] (-1,0) -- (17,0) node[right,scale=1.3]{$x$};
         \draw[black,thick,->] (-1,0) -- (-1,5) node [left, scale=1.3]{$t$};
        \shade [top color=Fuchsia, bottom color=cyan, opacity=0.5]
(6,0) rectangle (10,5);
\shade [top color=Fuchsia, bottom color=cyan, opacity=0.5]
(-1,0) rectangle (2,5);
\shade [top color=yellow, bottom color=red, opacity=0.8]
(2,0) rectangle (6,5);
\shade [top color=yellow, bottom color=red, opacity=0.8]
(10,0) rectangle (14,5);
\shade [top color=Fuchsia, bottom color=cyan, opacity=0.5]
(14,0) rectangle (17,5);
\draw[thick,black] (2,0)-- (2,5);
\draw[thick,black] (6,0)-- (6,5);
\draw[thick,black] (10,0)-- (10,5);
\draw[thick,black] (14,0)-- (14,5);

\draw[black,line width=2pt,->] (8,0)--(10.5,3);
\draw[black,line width=2pt,->] (8,0)--(5.5,3);
\draw [decorate,decoration={brace,amplitude=5pt,raise=1ex}]
  (5.5,3) -- (10.5,3) node[midway,above,yshift=1.4ex,scale=1.3]{$\tilde\rho_{\rm pure}$};
  
\draw[black,line width=2pt,->] (3.5,0)--(4.5,3) ;
\draw[black,line width=2pt,->] (3.5,0)--(2.5,3);
\draw [decorate,decoration={brace,amplitude=5pt,raise=1ex}]
  (2.5,3) -- (4.5,3) node[midway,above,yshift=1.4ex,scale=1.3]{${\rho}_{\rm pure}$};

\draw[black,line width=2pt,->] (13.5,0)--(14.5,3) ;
\draw[black,line width=2pt,->] (13.5,0)--(12.5,3);
\draw [decorate,decoration={brace,amplitude=5pt,raise=1ex}]
  (12.5,3) -- (14.5,3) node[midway,above,yshift=1.4ex,scale=1.3]{$\rho_{\rm mixed}$};

\node [scale=1.5] at (0,5.5){$\overline{A}$};
\node [scale=1.5] at (8,5.5){$\overline{A}$};
\node [scale=1.5] at (15.5,5.5){$\overline{A}$};
\node [scale=1.5] at (4,5.5){$A_1$};
\node [scale=1.5] at (12,5.5){$A_2$};
\node [scale=1.2] at (8,-0.5){$d$};
\node [scale=1.2] at (4,-0.5){$\ell_1$};
\node [scale=1.2] at (12,-0.5){$\ell_2$};
    \end{tikzpicture}
    \caption{Quasiparticle contributions to the reduced density matrix in a tripartite geometry. In the notation of \eqref{eq:densitymatrixdecomposition}, we see the two main contributions which will determine negativity and entanglement Hamiltonians, namely a pure part coming from pairs which are shared between the two subsystems, $\tilde{\rho}_{\rm pure}$ and a mixed part coming from pairs shared between the full $A$ and its complement, $\rho_{\rm mixed}$. Moreover, we have the fully pure contribution, $\rho_{\rm pure}$ which will have no contribution to the relevant entangling quantities.}
    \label{fig:qpp}
\end{figure}
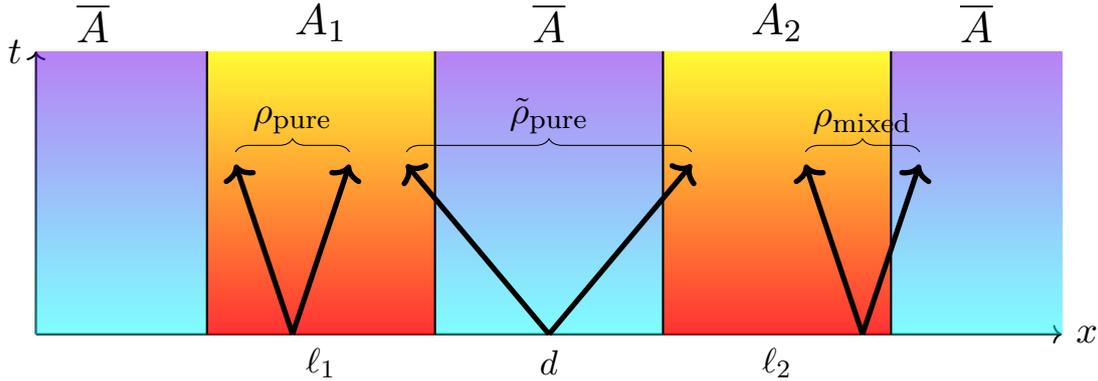

\subsection{Negativity and the tripartite geometry}
In this paper, we  aim to calculate the negativity Hamiltonian using the quasiparticle picture formalism. We thus take our subsystem of interest, $A$, to consist of two disjoint parts $A=A_1\cup A_2$ containing  $\ell_1$ and $\ell_2$ contiguous sites respectively, and lying $d$ sites apart, see figure~\ref{fig:qpp}. We take both $\ell_1$,$\ell_2$ and $d$ to be much larger than the fluid cell size, $\ell_1,\ell_2,d\gg\Delta$. Then, having in mind the decomposition~\eqref{eq:factorized} we can anticipate that in this case we have three types of contributions to $\rho_A(t)$ which factorize as, 
\begin{equation}
    \rho_A(t) = \rho_{\rm mixed}(t) \otimes \rho_{\rm pure}(t) \otimes \tilde{\rho}_{\rm pure}(t).
    \label{eq:densitymatrixdecomposition}
\end{equation}
Here, the $\rho_{\rm{pure}}(t)$ and $\rho_{\rm mixed}(t)$ are similar to case above: the pure part contains quasiparticles which are both in either $A_1$ or $A_2$, while the mixed part receives contributions from quasiparticles in $A$ whose partner was in $\bar{A}$ and has been traced over. As before, the former term does not contribute to entanglement, while the latter does. In what follows we shall ignore the former term since we are interested in the negativity Hamiltonian to which it does not contribute.  From the latter term, $\rho_{\rm mixed}(t)$, one can determine the entanglement Hamiltonian in the tripartite geometry as in ~\eqref{eq:ent_Ham}, however, it also contributes to the negativity Hamiltonian. The third term, $\tilde{\rho}_{\rm pure}(t)$, is new with respect to the bipartite case and is responsible for the entanglement between $A_1$ and $A_2$. It contains quasiparticle pairs which are both inside $A$ but for which one member of the pair is in $A_1$ and the other is in $A_2$. Thus, this term provides no contribution to the entanglement between $A$ and $\bar A$ and so does not contribute to $K_A(t)$, but is precisely the one which provides the non trivial contribution to the negativity and hence the negativity Hamiltonian. Throughout the remainder of the text we refer to $\rho_{\rm mixed}(t)$ as the mixed part and $\tilde{\rho}_{\rm pure}(t)$ as the pure part. 

From this picture, we are lead to following  decomposition of the negativity Hamiltonian into two distinct commuting terms
\begin{eqnarray}
   \mathcal{N}_A(t)=\mathcal{N}_{A,\rm m}(t)+\mathcal{N}_{A,\rm p}(t)~.
\end{eqnarray}
The first term comes form the mixed part
\begin{eqnarray}
    \rho_{\rm mixed}^{T_1}(t)=\frac{1}{\mathcal{Z_A}}e^{-\mathcal{N}_{A,\rm m}(t)}
\end{eqnarray}
which allows us to relate it directly to the tripartite entanglement Hamiltonian via $\mathcal{N}_{A,\rm m}(t)=K_A^{T_1}(t)$. The second term instead arises from
\begin{eqnarray}
     \tilde{\rho}^{T_1}_{\rm pure}(t)=e^{-\mathcal{N}_{A,\rm p}(t)}.
\end{eqnarray}
The same decomposition also applies to the fermionic negativity Hamiltonian with the appropriate replacements,
\begin{eqnarray}
       \mathcal{N}^{(f)}_A(t)=\mathcal{N}^{(f)}_{A,\rm m}(t)+\mathcal{N}^{(f)}_{A,\rm p}(t)~,
\end{eqnarray}
where 
\begin{eqnarray}
    \rho_{\rm mixed}^{R_1}(t)=\frac{1}{\mathcal{Z_A}}e^{-\mathcal{N}^{(f)}_{A,\rm m}(t)}~~,~~ ~~\tilde{\rho}^{R_1}_{\rm pure}(t)=e^{-\mathcal{N}^{(f)}_{A,\rm p}(t)}~.
\end{eqnarray}
We shall proceed by considering separately, $\mathcal{N}_A(t)$ and then $\mathcal{N}^{(f)}_A(t)$ starting in the next section with former. 

\section{Negativity Hamiltonian }
\label{sec:pure}
In this section we consider the negativity Hamiltonian $\mathcal{N}_A(t)$ and construct its contributions arising from the pure and mixed terms. We concentrate first on the squeezed state~\eqref{eq:squeezed} and following this, carry out the same procedure for the symmetric state~\eqref{eq:dimer}. 

\subsection{Partial transpose in the Hydrodynamic picture}
The first thing to understand is how to implement the partial transpose at the level of the fluid cell creation operators. The key observation is that the quasiparticle structure induces a very tight tensor product structure on the Hilbert space~\cite{travaglino2024quasiparticlepictureentanglementHamiltonians}. Hence, the state is effectively block diagonal in terms of the position $x$ and the momenta $k$, and we can perform the transposition separately on each $\rho_{x_0,k}(t)$. 
The transposition is, by its nature, basis dependent and the natural basis to choose is that of the quasiparticle number operator.  With this in mind, the definition of the partial transpose is
\begin{equation}
    \braket{ij | \rho_A^{T_1}| k l} = \braket{kj|\rho_A|i l}~,
\end{equation}
where the first index refers to $A_1$ while the second to $A_2$ and the indices label states in the chosen basis. In the occupation number basis, this can be expressed in terms of the creation and annihilation operators as,
\begin{equation}
    b_{x,k}^{T} = b^\dagger_{x,k},\hspace{0.5cm}  (b_{x,k}^\dagger)^{T} = b_{x,k},
\end{equation}
while the number operators themselves remain unvaried
\begin{equation}
    \hat{n}^T_{x,k}=(b_{x,k}^\dagger b_{x,k})^T = b_{x,k}^T (b_{x,k}^\dagger)^T = b_{x,k}^\dagger b_{x,k}=\hat{n}_{x,k}~.
\end{equation}
From this one can immediately obtain the contribution of the mixed part. Indeed, as explained above this is directly related to the entanglement Hamiltonian of the tripartite system via $ \mathcal{N}_{A,\rm m}(t)=K^{T_1}_A(t)$. Combining this with the fact that $\hat{n}^{T_1}_{x_t}(k)=\hat{n}_{x_t}(k)$ $\forall x_t$ we have that 
 \begin{eqnarray}
     \mathcal{N}_{A,\rm m}(t)&=&K_A(t)   
 \end{eqnarray}
with the right hand side being given by~\eqref{eq:ent_Ham_2} with $A = A_1 + A_2$. Thus the contribution of the mixed terms has a particularly simple and transparent form, since  the terms constituting the entanglement Hamiltonian are of the same form as those appearing in~\eqref{eq:ent_Ham_modes},
\begin{equation}
    \rho_{x_0,k}(t)= \frac{1}{1+e^{-\eta(k)}} e^{-\eta(k) \hat{n}_{x_t}(k)}.
\end{equation}
Including all the terms this leads to the full $\mathcal{N}_{A,m}$, which has the relatively simple form
\begin{equation}
    \mathcal{N}_{A,m} = \int \frac{dk}{2\pi} \int dx \chi_{\rm m}(x,k,t) \eta(k) \hat{n}_x(k),
\end{equation}
where we have introduced the counting function $\chi_{\rm m}(x,k,t)$, which selects only pairs which are shared between $A_1\cup A_2$ and its complement: this will be investigated explicitly in appendix \ref{appA}.
The expression can be easily inverted to real space by inserting the definition of the fluid cell operators, 
\begin{equation}
    \mathcal{N}_m = \int dx \int dz \mathcal{K}_m(x,z,t) c_x^\dagger c_{x+z}
    \label{eq:realspaceeh}
\end{equation}
where we have collected all momentum dependence in the kernel, 
\begin{equation}
    \mathcal{K}_m(x,z,t) = \int \frac{dk}{2\pi}\chi_{\rm m}(x,k,t) \eta(k) e^{ikz}.
\end{equation}
Note that this has the same structure of the bipartite entanglement Hamiltonian of \cite{rottoli2024}, up to the counting function which is different.

In contrast, the contribution of the pure terms is more intricate and requires a more in depth analysis. 
Recall that, by construction, $\tilde{\rho}_{\rm pure}(t)$ is made of quasiparticle pairs which are shared between $A_1$ and $A_2$. It can be written as
  \begin{eqnarray}
      \tilde\rho_{\rm pure}(t)=\prod_{k}\prod_{\{x_0 |x_t(-k)\in A_1,  x_t(k)\in A_2 \}}
     \rho_{x_0,k}(t).
  \end{eqnarray}
  Owing to the product structure, the partial transposition can be carried out on each term $\rho_{x_0,k}(t)$ separately. In particular, suppose that it is the left mover which is in $A_1$ and the right mover which is in $A_2$, i.e. $x_t(-k)\in A_1$. Then, taking the partial transposition on $A_1$ leads to
\begin{multline}
    \rho^{T_1}_{x_0,k}(t) = 
    n(k) \hat{n}_{x_t}(k)\hat{n}_{x_t}(-k)    + (1-n(k))(1- \hat{n}_{x_t}(k))(1-\hat{n}_{x_t}(-k)) \\ +  \sqrt{n(k)(1-n(k))}(e^{i\varphi_k}b_{x_t(k),k}^\dagger b_{x_t(-k),-k} +e^{-i\varphi_k} b_{x_t(-k),-k}^\dagger b_{x_t(k),k}).
    \label{eq:transposedrho} 
\end{multline}
where the only modification appears in the second line. Interestingly, despite the fact that the state breaks particle number symmetry, this density matrix now commutes with the particle number, indicative of the fact that the partial transposition does not respect global conservation laws~\cite{PhysRevA.98.032302}. 

To obtain $\mathcal{N}_{A,\rm p}(t)$ we should investigate the properties of $\rho^{T_1}_{x_0,k}(t)$ and to this end we denote the first and second lines of \eqref{eq:transposedrho} as $\mathcal{A}$  and $\mathcal{B}$ respectively.  It is then straightforward to check that, for any integer $p$, the first part obeys
\beqa\label{eq:A}
\mathcal{A}^p &=&  n(k)^p \hat{n}_{x_t}(k)\hat{n}_{x_t}(-k)   + (1-n(k))^p(1- \hat{n}_{x_t}(k))(1-\hat{n}_{x_t}(-k)),
\eeqa
whereas the second term displays a parity effect,
\begin{align}\label{eq:B}
\mathcal{B}^p &=& \begin{cases}
    \left[n(k)(1-n(k))\right]^{p/2} (e^{i\varphi_k}b_{x_t(k),k}^\dagger b_{x_t(-),-k} +e^{-i\varphi_k} b_{x_t(-k),-k}^\dagger b_{x_t(k),k})~& \text{  $p$ odd}\\
     \left[n(k)(1-n(k))\right]^{p/2} (\hat{n}_{x_t}(k) (1-\hat{n}_{x_t}(-k)) + \hat{n}_{x_t}(-k) (1-\hat{n}_{x_t}(k))) &\text{  $p$ even}. 
\end{cases}
\end{align}
In addition, using the fact that the first term, $\mathcal{A}$, is a sum of projectors onto particle number sectors in which $\mathcal{B}$ acts as $0$, one finds that the two terms are orthogonal,
\beq
\mathcal{A} \mathcal{B} = \mathcal{B} \mathcal{A}= 0~. 
\eeq
These properties then allow one to efficiently take powers of the transposed density matrix and find the negativity Hamiltonian. To do this, we note that the square of \eqref{eq:transposedrho} gives
\begin{equation}
    [\rho^{T_1}_{x_0,k}(t)]^2 = \frac{1}{(1+e^{-\eta(k)})^2} \exp\left(-\eta(k)(\hat{n}_{x_t}(k) + \hat{n}_{x_t}(-k))\right),
\end{equation}
which is reminiscent of the entanglement Hamiltonian~\eqref{eq:tracedrho}. To find $\mathcal{N}_{A,\rm p}(t)$, we need to find a square root of this expression. A natural ansatz for this is,
\begin{equation}
    \rho_{x_t,k}^{T_1} = \frac{1}{1+e^{-\eta(k)}}\exp\left\{-\frac{\eta(k)}{2}(\hat{n}_{x_t}(k) + \hat{n}_{x_t}(-k)) + i \frac{\pi}{2} \hat{O}_{x_t,k}\right\} 
    \label{eq:expnongaus}~,
\end{equation}
where the operator $\hat{O}_{x_t,k}$ is Hermitian and has an even integer spectrum such that the imaginary part only contributes a phase, 
\begin{equation}
    \hat{1} = \exp{i\pi \hat{O}_{x_t,k}}~.
\end{equation}
A nontrivial $\hat{O}_{x_t,k} $ is necessary to have a negative operator, which is what is expected from the transposed density matrix, in which negativity of the spectrum is precisely the signature of the Peres criterion. Using the explicit expression for~\eqref{eq:transposedrho} one can check that the correct choice is
\begin{eqnarray}\nonumber
    \hat{O}_{x_t,k} &=&\hat{n}_{x_t}(k)(1-\hat{n}_{x_t}(-k)) + \hat{n}_{x_t}(-k)(1-\hat{n}_{x_t}(k)) \\\label{eq:O_squeezed}&&-  e^{i\varphi_k} b_{x_t(k),k}^\dagger b_{x_t(-k),-k} - e^{-i\varphi_k}b_{x_t(-k),-k}^\dagger b_{x_t,k}.
\end{eqnarray}
This expression has several interesting features. First, $\hat{O}_{x_t,k}^2=2\hat{O}_{x_t,k}$ meaning that $\hat{O}_{k,t}$ has eigenvalues $0,\pm 2$ and moreover $[\hat{O}_{x_t,k},\hat{n}_{x_t}(k)+\hat{n}_{x_t}(-k)]=0$. These properties then guarantee that this operator contributes only to the sign of the eigenvalues but not their magnitude. Second, this operator contains terms which are quartic in fermions i.e. $\hat{n}_{x_t}(k)\hat{n}_{x_t}(-k)$ and so is not Gaussian. This implies therefore, that the negativity Hamiltonian of a quenched free fermion system is ``interacting''.  The partial transpose is not expected to preserve the Gaussianity of a state~\cite{Eisler_2015,Coser_2016}, however the fact that it breaks it via the introduction of a four-fermion interaction term is somewhat unexpected. 

We can now compute the contribution of the pure terms to the negativity Hamiltonian by summing up over all fluid cells and momentum contributions. Up to an additive constant this gives
\begin{eqnarray}\label{eq:neg_Ham_pure}
    \mathcal{N}_{A,\rm p}(t)&=&\sum_{k}\sum_{{\{x_0 |x_t(k)\in A_1,  x_t(-k)\in A_2 \}}} \frac{\eta(k)}{2}\left( \hat{n}_{x_t}(k)+\hat{n}_{x_t}(-k)\right)-i\frac{\pi}{2}\hat{O}_{x_t,k}.
\end{eqnarray}
Combining this with $\mathcal{N}_{A,\rm m}(t)$ we obtain the full expression of $\mathcal{N}_{A}(t)$. We can then perform the inverse Fourier transform to bring this back to an expression in real space. However, non-Gaussianity makes this quite tricky and also prevents us from checking this result against exact numerics. Therefore, we will only study the real space version of the fermionic negativity Hamiltonian. Nevertheless, as we will show later, this form reproduces all known results for the negativities and their symmetry decomposition. 

\subsection{Symmetric state}
We now repeat the same analysis for the symmetric state which preserves the $U(1)$ particle number symmetry. As before, we consider separately the two contributions, $\mathcal{N}_{A,\rm m}(t)$ and $\mathcal{N}_{A,\rm p}(t)$. It can be noted, however, that the entanglement Hamiltonians for both types of states~\eqref{eq:squeezed} and~\eqref{eq:dimer} depend only on $n(k)$ and so are in fact the same. Thus we immediately obtain that $\mathcal{N}_{A,\rm m}(t)=K_A(t)$. The difference between the two states is instead revealed by the contribution of the pure terms. 

Consider, a single $\rho_{x_0,k}(t)$, given by~\eqref{eq:symm_presrve_pure}, in which one quasiparticle, the left mover, is in $A_1$ and the other is in $A_2$ i.e. $x_t(k-\pi)\in A_1,~x_t(k)\in A_2$ . Upon partial transposition this we find,
\begin{multline}
    \rho^{T_1}_{x_0,k}(t) = 
    n(k) \hat{n}_{x_t}(k)(1-\hat{n}_{x_t}(k-\pi) )   + (1-n(k))(1- \hat{n}_{x_t}(k))\hat{n}_{x_t}(k-\pi) \\+  \sqrt{n(k)(1-n(k))}(e^{i\varphi_k}b_{x_t(k),k}^\dagger b^\dag_{x_t(k-\pi),k-\pi} +e^{-i\varphi_k} b_{x_t(k-\pi),k-\pi} b_{x_t(k),k})~.\label{eq:transposedrho_dimer}
\end{multline}
Here, one sees that the first line is unaffected by the transposition while the  second is. In particular, it now contains terms which break the particle number symmetry. Once again this is not a surprise since partial transposition does not respect particle number~\cite{PhysRevA.98.032302}, but exchanges it for symmetry with respect to $N_{A_2}-N_{A_1}$ i.e. the imbalance in particle number between the two parts of $A$. Inspecting~\eqref{eq:transposedrho_dimer} one can see that this is indeed preserved since, in the second line, particles are either created or destroyed pairwise in $A_1$ and $A_2$. 

Proceeding as before, one can then express this in exponential form which is analogous to~\eqref{eq:expnongaus},
\begin{equation}
    \rho_{x_t,k}^{T_1} = \frac{1}{1+e^{-\eta(k)}}\exp\left\{-\frac{\eta(k)}{2}(\hat{n}_{x_t}(k) + (1-\hat{n}_{x_t}(k-\pi))) + i \frac{\pi}{2} \hat{O}_{x_t,k}\right\} 
    \label{eq:expnongaus}~,
\end{equation}
but with a different operator $\hat{O}_{x_t,k}$ which is given by 
\begin{eqnarray}\nonumber
    \hat{O}_{x_t,k} &=&\hat{n}_{x_t}(k)\hat{n}_{x_t}(k-\pi) +(1-\hat{n}_{x_t}(k)) (1- \hat{n}_{x_t}(k-\pi)) \\ \label{eq:O_dimer}&&-  e^{i\varphi_k} b_{x_t(k),k}^\dagger b^\dagger_{x_t(k-\pi),k-\pi} - e^{-i\varphi_k}b_{x_t(k-\pi),k-\pi}^\dagger b^\dagger_{x_t,k}.
\end{eqnarray}
which has similar spectral and algebraic properties to the squeezed state version. Thus we find that $\mathcal{N}_{A,\rm p}(t)$ has the same form as~\eqref{eq:neg_Ham_pure} but with the application of an appropriate particle-hole exchange. In particular, it is non-Gaussian, containing quartic fermion terms.

\section{Fermionic Negativity Hamiltonian}\label{sec:fermionic}
We now turn our attention to the fermionic negativity Hamiltonian. We proceed in an analogous fashion and consider first the squeezed state and then the symmetric state.

\subsection{Partial time reversal in the Hydrodynamic picture}
Our first task is to understand how to implement the partial time reversal operation in our hydrodynamic framework. Just like for the partial transpose, this task is eased by the tensor product structure of the modes and fluid cells and so we can apply it to each separately. The basic definition of fermionic transposition is obtained from the fermionic coherent states, and is defined such that 
\begin{equation}
    \braket{\{\xi\}_1, \{\xi\}_2 | \rho_A^{R_1} | \{\overline{\xi}\}_1, \{\overline{\xi}\}_2} = \braket{\{i\overline{\xi}\}_1, \{\xi\}_2 | \rho_A | \{i\xi\}_1, \{\overline{\xi}\}}
    \label{eq:fermionictranspose}
\end{equation}
Where $\{\xi\}_{1,2}$ represent sets of coherent states in the subspaces $A_1$ and $A_2$ respectively. In the hydrodynamic effective description, there will be $(x,k)$ dependent coherent states, related to the fluid cell operators $b_{x,k},b_{x,k}^\dagger$ via the usual definition, 
\begin{equation}
    \ket{\xi_{x,k}} = e^{-\xi b_{x,k}^\dagger} \ket{0}, \hspace{0.5cm}\bra{\overline{\xi}_{x,k}} = \bra{0}e^{-b_{x,k} \overline{\xi}}~,
\end{equation}
where $\xi,\overline{\xi}$ are Grassmann numbers. 
Using the properties of coherent states, it is not difficult to convince oneself that this operation leaves the number operator invariant and so  $\hat{n}^{R_1}_{x_t}(k)=\hat{n}_{x_t}(k)$ $\forall x_t$. Thus we are lead to the same relation between the entanglement Hamiltonian and the contribution of the mixed terms to the fermionic negativity Hamiltonian,
\beqa \label{eq:fermionmixedham}
   \mathcal{N}^{(f)}_{A,\rm m}(t)&=&K_A(t)   ~.
\eeqa
The difference between the entanglement and fermionic negativity Hamiltonians once again lies in the contribution of the pure part.

We take the local density matrix for the squeezed state~\eqref{eq:squeezed} and consider the case in which the left mover is inside $A_1$ and the right mover is inside $A_2$. Upon using some standard identities of fermionic coherent states we find that the partial time reversal~\eqref{eq:fermionictranspose} operation returns
\begin{multline}
    \rho^{R_1}_{x_0,k}(t) =
    n(k) \hat{n}_{x_t}(k)\hat{n}_{x_t}(-k)    + (1-n(k))(1- \hat{n}_{x_t}(k))(1-\hat{n}_{x_t}(-k)) \\+  i\sqrt{n(k)(1-n(k))}(e^{i\varphi_k}b_{x_t(k),k}^\dagger b_{x_t(-k),-k} +e^{-i\varphi_k} b_{x_t(-k),-k}^\dagger b_{x_t(k),k}).
    \label{eq:fermiontransposedrho}
\end{multline}
Note that the difference with respect to the previous result~\eqref{eq:transposedrho} is minimal, and amounts solely the $i$ factor in front of the off-diagonal part. This makes the operator non-Hermitian, as is standard for the partial time reversal, but oppositely to the partial transpose which is always Hermitian.
In order to exponentiate this, we consider, instead of its square as in the previous section, the product with its conjugate,
\begin{equation} \rho_{x_0,k}^{R_1}(t)\left[\rho_{x_0,k}(t)^{R_1}\right]^\dagger= \frac{1}{(1+e^{-\eta(k)})^2}\exp\left(-\eta(\hat{n}_{x_t}(k)+\hat{n}_{x_0}(-k))\right).
\end{equation}
This suggests the following ansatz for $\rho_{x_0,k}^{R_1}(t)$, 
\begin{equation}\label{eq:fermionictransposedrho_squeezed}
    \rho_{x_0,k}^{R_1}(t) = \frac{1}{1+e^{-\eta(k)}}\exp\left(-\frac{\eta(k)}{2}(\hat{n}_{x_t}(k)+\hat{n}_{x_t}(-k)) + i\frac{\pi}{2}\hat{O}^{(f)}_{x_t,k}\right)
\end{equation}
which is similar to the previous case, but where now the only requirement on $\hat{O}^{(f)}_{x_t,k}$ is its Hermiticity, with no additional conditions on the spectrum. By expanding out the exponential and comparing to~\eqref{eq:fermiontransposedrho} one finds that the correct choice is 
\begin{equation}\label{eq:O_squeezed_fermion}
    \hat{O}^{(f)}_{x_t,k} = e^{i\varphi_k} b^\dagger_{x_t(k),k}b_{x_t(-k),-k} +e^{-i\varphi_k} b_{x_t(-k),-k}^\dagger b_{{x_t(k),k}} .
\end{equation}
Notably, this operator is quadratic in fermions, in stark contrast to~\eqref{eq:O_squeezed}. As a result, upon inserting this into our previous expression, we find that $ \rho_{x_0,k}^{R_1}(t)$ is Gaussian, as expected. Up to an additive constant, the contribution of the pure terms to the fermionic negativity Hamiltonian is, 
\begin{eqnarray}\label{eq:neg_Ham_pure}
    \mathcal{N}^{(f)}_{A,\rm p}(t)&=&\sum_{k}\sum_{{\{x_0 |x_t(k)\in A_1,  x_t(-k)\in A_2 \}}} \frac{\eta(k)}{2}\left( \hat{n}_{x_t}(k)+\hat{n}_{x_t}(-k)\right)-i\frac{\pi}{2}\hat{O}^{(f)}_{x_t,k}
\end{eqnarray}
This can be combined with $\mathcal{N}^{(f)}_{A,\rm m}(t)$ to obtain the full expression for $\mathcal{N}^{(f)}_A(t)$ which is quadratic.  It is remarkable that the small change in~\eqref{eq:fermiontransposedrho} compared to~\eqref{eq:transposedrho} results in a quadratic fermionic negativity Hamiltonian instead of an interacting one. 

As discussed in the introduction, the partial transpose applied to a Gaussian state leads to a sum of two Gaussian states, one of which is equivalent to the partially time reversed state. We can see this in our formalism by noting that 
\begin{eqnarray}
    \rho_{x_0,k}^{T_1}(t)=\frac{1-i}{2}\rho_{x_0,k}^{R_1}(t)+\frac{1+i}{2}[\rho_{x_0,k}^{R_1}(t)]^\dagger
\end{eqnarray}
in agreement with the result of~\cite{Shapourian2017}. Note that this applies on each block separately, not on the full density matrix which is a product over many such terms. 

\subsection{Symmetric state}
We can repeat the same procedure for the symmetry preserving, symmetric states. Here we need only examine the contribution of the pure terms, since, as above $\mathcal{N}_{A,\rm m}(t)=K_A(t)$, which is identical for the two types of states. Thus we consider a single $\rho_{x_0,k}(t)$, given by~\eqref{eq:symm_presrve_pure} with the left moving quasiparticle being contained in $A_1$ and the right moving counterpart in $A_2$. Upon carrying out the partial time reversal, we find 
\begin{multline}
    \rho^{R_1}_{x_0,k}(t) = 
    n(k) \hat{n}_{x_t}(k)(1-\hat{n}_{x_t}(k-\pi) )   + (1-n(k))(1- \hat{n}_{x_t}(k))\hat{n}_{x_t}(k-\pi) \\ +  i\sqrt{n(k)(1-n(k))}(e^{i\varphi_k}b_{x_t(k),k}^\dagger b^\dag_{x_t(k-\pi),k-\pi} +e^{-i\varphi_k} b_{x_t(k-\pi),k-\pi} b_{x_t(k),k})~.\label{eq:fermiontransposedrho_dimer}
\end{multline}
As was the case for the squeezed state, the only difference with the partial transpose is the appearance of the $i$ in the second line.  This modification, is crucial in preserving the Gaussianity of the state which has a similar form as as~\eqref{eq:fermionictransposedrho_squeezed},
\begin{equation}
      \rho_{x_0,k}^{R_1}(t) = \frac{1}{1+e^{-\eta(k)}}\exp\left(-\frac{\eta(k)}{2}(\hat{n}_{x_t}(k)+(1-\hat{n}_{x_t}(k-\pi))) + i\frac{\pi}{2}\hat{O}^{(f)}_{x_t,k}\right)
\end{equation}
together with the replacement 
\begin{eqnarray}\label{eq:O_dimer_fermion}
    \hat{O}^{(f)}_{x_t,k} = e^{i\varphi_k} b^\dagger_{x_t(k),k}b^\dagger_{x_t(k-\pi),k-\pi} +e^{-i\varphi_k} b_{x_t(k-\pi),k-\pi} b_{{x_t(k),k}} .
\end{eqnarray}
The full fermionic negativity Hamiltonian follows by summing over the different contributions \textit{\'a la}~\eqref{eq:neg_Ham_pure} and combining it with $K_A(t)$. Note that $\mathcal{N}^{(f)}_{A}(t)$ does not preserve particle number symmetry due to the explicit form of~\eqref{eq:O_dimer_fermion} but does respect the particle number imbalance, $N_{A_2}-N_{A_1}$ as would be expected~\cite{PhysRevA.98.032302}. We will analyse the charge imbalance in detail in section \ref{sec:charge_imbalance}.
Finally, note that the exponent can be diagonalized by a simple transformation,
\begin{equation}
     b_{\pm}^\dagger = \frac{b_{x_t(k),k}^\dagger \pm b_{x_t(k-\pi),k-\pi}}{\sqrt{2}}
\end{equation}
which leads to the diagonal form 
\begin{equation}
      \rho_{x_0,k}^{R_1}(t) \propto \exp\left(-\left(\frac{\eta(k)}{2}-\frac{i\pi}{2}\right) \hat{n}_{+} -\left(\frac{\eta(k)}{2}+\frac{i\pi}{2}\right) \hat{n}_{-} \right)~.
\end{equation}
which immediately shows that the negativity Hamiltonian is composed of pairing excitations of imaginary energy. 
This formulation allows one to determine even more readily the negativities from $\mathcal{N}_A^{(f)}(t)$.

\subsection{Real space form}
\label{sec:realspace_2}
As with the entanglement Hamiltonian, it is often desirable to present $\mathcal{N}^{(f)}_A(t)$ in real space in order to understand its spatial structure. The full expression for the fermionic negativity Hamiltonian is conveniently written inserting a new counting function,
\begin{equation}
\label{eq:nhrewriting}
    \mathcal{N}^{(f)}_A(t) = K_A(t)+ \int_{k>0} \frac{{\rm d}k}{4\pi} \int {\rm d}x  \,\chi_{\rm p}(x,k,t) \left(\eta(k)(\hat{n}_{x,k} + \hat{n}_{x-2v_k t,-k}) - i \pi \hat{O}^{(f)}_{x_t,k}\right).
\end{equation}
where $\chi_{\rm p}(x,k,t)$ selects pairs which are shared between $A_1$ and $A_2$, and we have changed the sums to integrals for simplicity. In the case of the symmetric states it would be necessary to substitute $\hat{n}_{x-2v_k t,-k}(-k) \to (1- \hat{n}_{x-2v_k t,k-\pi}(k-\pi))$ and to use the corresponding $\hat{O}^{(f)}_{x_t,k}$ operator.

This Hamiltonian is purely quadratic and remarkably can be obtained from $\mathcal{N}_{A}(t)$ by simple removing the non-quadratic terms. It is convenient to express it in a way which highlights the different terms which constitute it,
\begin{equation}
      \mathcal{N}^{(f)}_A(t) = K_A(t) + \mathcal{N}_{A,d}^{(f)} (t) + \mathcal{N}_{A,nd}^{(f)} (t).
\end{equation}
The first term is just the entanglement Hamiltonian while the second is a diagonal term $\mathcal{N}_{A,d}^{(f)} (t)$ which arises from the terms of \eqref{eq:nhrewriting} containing only the number operators. This can be immediately inverted back to real space by repeating the same analysis valid for the entanglement Hamiltonian, leading to 
\begin{equation}
\label{eq:diagpart}
   \mathcal{N}_{A,d}^{(f)} (t)=\int_{x\in A_2} {\rm d}x \int {\rm d}z \mathcal{K}_+(x,z,t) c^\dagger_x c_{x+z} + \int_{x\in A_1} {\rm d}x \int  {\rm d}z \,\mathcal{K}_-(x,z,t)c^\dagger_{x}c_{x+z} 
\end{equation}
where we have introduced the kernels
\beqa
    \mathcal{K}_+(x,z,t) &=& \int_{k>0} \frac{ {\rm d}k}{4\pi} \chi_{\rm p}(x,k,t) e^{ikz}\eta(k), \\
    \mathcal{K}_-(x,z,t) &=& \int_{k>0} \frac{ {\rm d}k}{4\pi} \chi_{\rm p}(x-2|v_k|t,k,t) e^{-ikz}\eta(k).
\eeqa
The last term is,  on the other hand, off-diagonal and arises from the inversion of $\hat{O}^{(f)}_{x_t,k}$. In general, the structure is simply
\begin{equation}
    \mathcal{N}^{(f)}_{A,nd}(t) = -\frac{i\pi}{2} \int_{x\in A_2}  {\rm d}x \int_{y\in A_1}  {\rm d}y \left(\mathcal{K}_{nd}(x,y,t)  c_x^\dagger c_y + h.c.\right). 
\end{equation}
However, specifying the kernel is a subtle task. In fact, the coarse graining cannot be inverted in this case, for the clear reason that $\hat{O}^{(f)}_{x_t,k}$ couples two single modes given a momentum $k,-k$.  This is quite different from the diagonal part, in which the sum over $k$ guarantees an averaging of the relevant modes, thus allowing to invert effectively the coarse graining. However, we can still note one significant property of this term: since $b^\dagger_{x,k} b_{x-2v_kt,k-\pi} + {\rm h.c.}$ is a square root of an idempotent operator (in particular, not positive but Hermitian), its eigenvalues are $0,\pm1$. Therefore  $ \mathcal{N}^{(f)}_{A,nd}(t)$ needs to have eigenvalues $0,\pm \frac{i\pi}{2}$. This feature is indeed confirmed by exact numerics, as will be shown in section \ref{sec:numerics}, and guarantees that this off-diagonal term contributes very simple factors $\pm i$ to the full fermionic transpose $\rho^{R_1}$. 

In the case of the symmetric state, the generalization is straightforward. The presence of $(1-\hat{n}_{x-2v_kt,k-\pi})$ leads to the simple modification 
\begin{equation}
\label{eq:diagonalsymmetric}
     \mathcal{N}_{A,d}^{(f)} (t)=\int_{x\in A_2}  {\rm d}x\int  {\rm d}z \mathcal{K}_+(x,z,t) c^\dagger_x c_{x+z} + \int_{x\in A_1} {\rm d}x \int  {\rm d}z \mathcal{K}_-(x,z,t)c_{x+z}c_{x}^\dagger,   
\end{equation}
where the kernels are slightly different from the case above,
\beqa
    \mathcal{K}_+(x,z,t) &=& \int_{k>0} \frac{ {\rm d}k}{4\pi} \chi_{\rm p}(x,k,t) e^{ikz}\eta(k), \\
    \mathcal{K}_-(x,z,t) &=& \int_{k>0} \frac{ {\rm d}k}{4\pi} \chi_{\rm p}(x-2|v_k|t,k,t) (-1)^z e^{ikz}\eta(k).
    \label{eq:kernelminusdimer}
\eeqa
while the non-diagonal part is immediately modified to
\begin{equation}
\label{eq:offdiag_symmetric}
    \mathcal{N}^{(f)}_{A,nd}(t) = -\frac{i\pi}{2} \int_{x\in A_2}  {\rm d}x \int_{y\in A_1}  {\rm d}y \left(\mathcal{K}_{nd}(x,y,t)  c_x^\dagger c^\dagger_y + h.c.\right).
\end{equation}
Again, although the kernel cannot be fixed by a simple inversion, this has single particle eigenvalues $0,\pm \frac{i\pi}{2}$, and this will be the main thing that will be checked numerically from an exact approach. Thus we have the rather interesting occurrence of non-local pair-hopping terms in the operatorial characterization of mixed state entanglement in symmetric states. This structure is nevertheless not too surprising, as it is well known that fermionic transposition changes the symmetry of the original state,
\begin{equation}
    [\rho, \hat{N}_1+\hat{N}_2] = 0 \iff  [\rho^{R_1}, \hat{N_2}-\hat{N}_1]=0.
\end{equation}
It is clear that a pairing term of the form \eqref{eq:offdiag_symmetric} is precisely the kind of term which would be expected to implement such kind of property. Although the explicit form of the kernel cannot be fixed analytically by implementing the inversion, it is possible to extrapolate its main features from the numerical analysis. This will be discussed in section \ref{sec:numerics}.

As a final comment, we observe that the contribution coming from the pure part disappears in the long time limit, since as $t \to \infty$ there can be no pairs which are shared between $A_1$ and $A_2$. Therefore in this limit only the mixed part contributes, and this in turn implies that
\begin{equation}
    \lim_{t \to \infty} \mathcal{N}_A(t) =  \lim_{t \to \infty} \mathcal{N}_A^{(f)} (t)= \lim_{t \to \infty}K_A(t)~.
\end{equation}
Since the entanglement Hamiltonian saturates to a Generalized Gibbse Ensemble (GGE) (see appendix \ref{appA}), as 
\begin{equation}
    \lim_{t \to \infty} K_A(t) = \int_{A_1}  {\rm d}x \int \frac{dk}{2\pi}\eta(k) \hat{n}_x(k) + \int_{A_2}  {\rm d}x \int \frac{dk}{2\pi}\eta(k) \hat{n}_x(k), 
    \label{eq:gge}
\end{equation}
this implies that the negativity Hamiltonian will itself tend to a GGE in the long time limit. This reflects the intuition that at long times the correlations within the system are dominated by classical contributions, which are fully taken into account by the entanglement Hamiltonian, and  genuinely negativity-related effects disappear completely.

\section{Analytic checks}
\label{sec:checks}
In this section we verify our expressions for the negativity and fermionic negativity Hamiltonians by showing that they reproduce previous calculations for the quench dynamics of the negativity~\cite{Alba_2019,Murciano2022} and imbalance resolved negativity~\cite{Parez_2022}.

\subsection{ R\'enyi negativities}
To begin, we can note that because of the product structure, the calculation of the R\'enyi negativities and their fermionic versions, splits into a contribution from the mixed terms and the pure terms. The former is easiest to calculate and using~\eqref{eq:ent_Ham_modes} one finds that 
\begin{eqnarray}
 \Tr[(\rho^{T_1}_{\rm mixed}(t))^\alpha]&=&\prod_{k}\prod_{\{x_0|x_t(-k)\in A ~\& ~x_t(k)\notin A\}}\Tr[(\rho_{x_0,k}(t))^\alpha]\\\label{eq:mixed_nega}
 &=&\prod_{k}\prod_{\{x_0|x_t(-k)\in A ~\& ~x_t(k)\notin A\}}\left[n(k)^\alpha+(1-n(k))^\alpha\right]\\
 &=&\Tr[(\rho^{R_1}_{\rm mixed}(t))^\alpha]=e^{(1-\alpha)S^\alpha_A(t)},
\end{eqnarray}
where in the last line, we have used the fact that $\mathcal{N}_{A,\rm m}(t)=\mathcal{N}^{(f)}_{A,\rm m}(t)=K_A(t)$ and introduced the R\'enyi entanglement entropy which is defined in the usual way $S^\alpha_A(t)=\frac{1}{1-\alpha}\log \Tr[\rho_A(t)^\alpha]$. Thus, the mixed contribution to the R\'enyi negativity  is the same as the fermionic version and both are simply related to the R\'enyi entanglement entropy between $A$ and $\bar A$. 

The pure terms require more work to evaluate, but this can be achieved using the properties given in \eqref{eq:A} and \eqref{eq:B} (alternatively, the contributions can be inferred by evaluating the trace over the basis which diagonalizes simultaneously both term appearing in the exponentiated form of the pair density matrix, which can be easily inferred in the $b_{x,k}$ basis). 
The results are the same for both the squeezed and symmetric states but depend on whether $\alpha$ is even or odd. For $\alpha_o$ being an odd integer
\begin{eqnarray}
 \Tr[(\tilde\rho^{T_1}_{\rm pure}(t))^{\alpha_o}]&=&\prod_{k}\prod_{\{x_0|x_t(-k)\in A_1 ~\& ~x_t(k)\in A_2\}}\Tr[(\rho^{T_1}_{x_0,k}(t))^{\alpha_o}]\\
 &=&\prod_{k}\prod_{\{x_0|x_t(-k)\in A_1 ~\& ~x_t(k)\in A_2\}}\left[n(k)^{\alpha_o}+(1-n(k))^{\alpha_o}\right]
\end{eqnarray}
which has a similar form to the R\'enyi entanglement entropy, c.f.~\eqref{eq:mixed_nega}. If, however,  $\alpha_e$ is even we get 
\begin{eqnarray}
 \Tr[(\tilde\rho^{T_1}_{\rm pure}(t))^{\alpha_e}]&=&\prod_{k}\prod_{\{x_0|x_t(-k)\in A_1 ~\& ~x_t(k)\in A_2\}}\Tr[(\rho^{T_1}_{x_0,k}(t))^{\alpha_e}]\\
 &=&\prod_{k}\prod_{\{x_0|x_t(-k)\in A_1 ~\& ~x_t(k)\in A_2\}}\left[n(k)^{\alpha_e/2}+(1-n(k))^{\alpha_e/2}\right]^2.
\end{eqnarray}
Here we see that this is related, instead, to the R\'enyi entropy with index $\alpha_e/2$.  

The same calculation can also be carried out to find the contribution of the pure terms to the fermionic negativity. Using the expressions~\eqref{eq:fermiontransposedrho} and \eqref{eq:fermiontransposedrho_dimer}, one arrives at the remarkable result, that for both types of states 
\begin{eqnarray}
   \Tr[(\tilde\rho^{T_1}_{\rm pure}(t))^{\alpha_o}] &=&\Tr[(\tilde\rho^{R_1}_{\rm pure}(t)[\tilde\rho^{R_1}_{\rm pure}(t)]^\dagger)^{\frac{\alpha_o-1}{2}}\tilde\rho^{R_1}_{\rm pure}(t)]~, \\
    \Tr[(\tilde\rho^{T_1}_{\rm pure}(t))^{\alpha_e}]&=& \Tr[(\tilde\rho^{R_1}_{\rm pure}(t)[\tilde\rho^{R_1}_{\rm pure}(t)]^\dagger)^{\frac{\alpha_e}{2}}]~.
\end{eqnarray}
This then directly implies that in the hydrodynamic regime, for the cases considered,
\begin{eqnarray}
\mathcal{E}_\alpha^{(f)}(t)=\mathcal{E}_\alpha(t)~~\forall \alpha~.
\end{eqnarray}
Therefore, within the quasiparticle picture and for integrable quenches in free fermionic models,  the negativity and the fermionic negativity are identical. This is despite the fact that the associated negativity Hamiltonians are considerably different with one being Gaussian and the other not.  

Using the explicit form of the counting functions given in Appendix~\ref{appA} and dropping the, now unnecessary, superscript $(f)$ 
we obtain the full quasiparticle prediction for the negativity
\begin{eqnarray}\nonumber
     \mathcal{E}_{\alpha}(t) &=&\int\frac{{\rm d}k}{2\pi}\left[\max(|v_k|t,d/2)-2\max(|v_k|t,(\ell+d)/2) + \max(|v_k|t,\ell+d/2)\right] \tilde{s}_\alpha(k)\\&&+(1-\alpha)S^\alpha_A(t)
     \label{eq:renyi_negativity}
\end{eqnarray}
where $\ell_1=\ell_2=\ell$ and
\begin{equation}
    \tilde{s}_\alpha(k) = \begin{cases}
        \log\left( n(k)^\alpha   + (1-n(k))^\alpha \right)~&~\text{ for $\alpha$ odd}\\
       2\log\left( n(k)^{\alpha/2}   + (1-n(k))^{\alpha/2}\right) &~\text{ for $\alpha$ even}
    \end{cases}
    \label{eq:varepsilon}
\end{equation}
with the explicit form of $S_A^\alpha(t)$ is given in~Appendix \ref{appA}. Moreover, the ratios $R_\alpha$ which display some universal properties are obtained as
\begin{eqnarray}
   \log R_\alpha(t)= \mathcal{E}_\alpha(t)+(\alpha-1)S_{\alpha}(t)~.\label{eq:ratios}
\end{eqnarray}
as obvious from their definition.
The above expressions reproduce the results of \cite{Murciano2022} which had been conjectured using CFT and verified numerically. 
It is important to highlight that Eq.~\eqref{eq:renyi_negativity} is quite general, being valid for both the squeezed states and the symmetric states. Thus, we believe it to be the general result for any state exhibiting a pair structure. 
In the replica limit we then find that the negativity itself is 
\begin{eqnarray}\nonumber
     \mathcal{E}(t) &=&\int\frac{{\rm d}k}{2\pi}\left[\max(|v_k|t,d/2)-2\max(|v_k|t,(\ell+d)/2) + \max(|v_k|t,\ell+d/2)\right] \tilde{s}_1(k)
     \label{eq:negativity}
\end{eqnarray}
in agreement with \cite{Alba_2019}. Once again, this expression is valid for both types of negativity and states which we have considered. 

\begin{figure}[h!]
    \centering
    \includegraphics[width=0.9\linewidth]{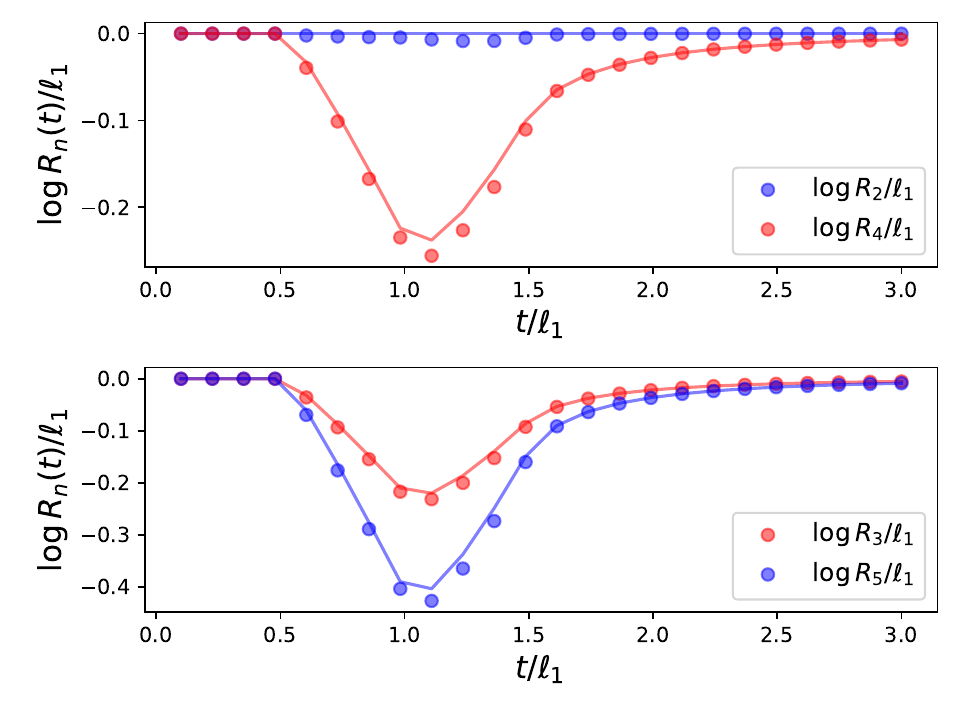}
    \caption{Time dependence of $\log R_n$ for two subsystems of length $\ell=800$ placed at distance $d=800$, in a quench from a dimer state. The symbols are the exact solution, which is obtained by evaluating \eqref{eq:renyinegativitiesdefinition} using the eigenvalues of the exact negativity Hamiltonian, computed with the methods discussed in section \ref{sec:numerics}, while the continuous line represents the theoretical prediction, which corresponds to the first line of \eqref{eq:renyi_negativity}. As expected, $\log R_2(t) $ is always zero, as $\tilde s_2(k) = 0$. }
    \label{fig:renyinegativities}
\end{figure}
In figure \ref{fig:renyinegativities} the theoretical prediction for the ratios \eqref{eq:ratios} is shown in comparison with exact numerical evaluations, which allow to write the R\'enyi negativities in terms of the eigenvalues of the negativity Hamiltonian for a quench from a dimer state (see section \ref{sec:numerics} for details). The figure exhibits perfect agreement between the quasiparticle result and the exact solution, therefore confirming our results. Small deviations from the theoretical result are mostly accounted for by the logarithmic correction which normally appears in the quasiparticle picture context, and by a cutoff on the spectrum of the numerical negativity Hamiltonian, which will be discussed in the following.

\subsection{Symmetry resolved negativities}
\label{sec:charge_imbalance}
In the previous section, we have shown that our negativity Hamiltonians can be used to reproduce previous calculations concerning the quench dynamics of the R\'enyi negativities. In doing so, we saw that both types of states, symmetry breaking and symmetry preserving, lead to the same final expression in terms of their occupation function. 
This was despite the quite different form of their respective negativity Hamiltonians and in particular the form of the operators $\hat{O}_{x_t,k}$, $\hat{O}^{(f)}_{x_t,k}$ given in ~\eqref{eq:O_squeezed}, ~\eqref{eq:O_dimer},~\eqref{eq:O_squeezed_fermion} and~\eqref{eq:O_dimer_fermion}. To expose the difference, one can compute the particle imbalance resolved negativities~\cite{PhysRevA.98.032302}. This has been studied for quenches in free fermions previously in \cite{Parez_2022} and in this section we reproduce these results as a further check on our expressions for $\mathcal{N}_A(t)$ and $\mathcal{N}^{(f)}_A(t)$.

Considering symmetric states of the form \eqref{eq:dimer}, $\rho_A(t)$ commutes with the total number operator $N_A$. However, this property is not preserved by the partial transpose or time reversal: it was shown instead that  $\rho^{T_1}_A$ actually commutes with $Q_A=N_{A_2} - N_{A_1}$~\cite{PhysRevA.98.032302} known as particle number imbalance. The imbalance resolved negativity seeks to understand how the negativity is split between the different charge sectors of the imbalance. The central quantity in this approach are the charged moments, defined as 
\begin{eqnarray}
N_\alpha(\lambda,t)&=&\Tr[e^{i\lambda Q_{A}}(\rho^{T_1}_A(t))^\alpha]
\end{eqnarray}
with a similar definition for the fermionic version, $N_\alpha^{(f)}(\lambda,t)$.  It is these objects which we obtain using our results. 

Since the number operator is naturally split in the quasiparticle space i.e. $Q_A=\sum_{x\in A_2}\hat{n}_{x,k}-\sum_{x\in A_1}\hat{n}_{x,k}$, we can study the effect of the symmetry resolution at the level of each quasiparticle pair. As with previous calculations, the contribution of the mixed terms is quite straightforward, but we must consider the pairs in $A_1$ and $A_2$ separately. We find 
\begin{eqnarray}\nonumber
 \Tr[e^{i\lambda Q_A}(\rho^{T_1}_{\rm mixed}(t))^\alpha]
 &=&\prod_{k}\prod_{\{x_0|x_t(-k)\in A_1 ~\& ~x_t(k)\notin A\}}\Tr[e^{-i\lambda \hat{n}_{x_t}(k)}(\rho_{x_0,k}(t))^\alpha]\\
 &&\times \prod_{k}\prod_{\{x_0|x_t(-k)\in A_2 ~\& ~x_t(k)\notin A\}}\hspace{-10pt}\Tr[e^{i\lambda \hat{n}_{x_t}(k)}(\rho_{x_0,k}(t))^\alpha]\\\nonumber
 &=&\prod_{k}\prod_{\{x_0|x_t(-k)\in A_1 ~\& ~x_t(k)\notin A\}}\hspace{-20pt}\big[e^{-i\lambda}n(k)^\alpha+(1-n(k))^\alpha\big]\\ &&
 \times\prod_{k}\prod_{\{x_0|x_t(-k)\in A_2 ~\& ~x_t(k)\notin A\}}\hspace{-30pt}\big[e^{i\lambda}n(k)^\alpha+(1-n(k))^\alpha\big]\\
 &=&\Tr[e^{i\lambda Q_A}(\rho^{R_1}_{\rm mixed}(t))^\alpha]~.
\end{eqnarray}
In the last line, we have used the fact that the partial transpose and partial time reversal have the same effect on $\rho_{\rm mixed}(t)$.  Considering then the contributions to the pure terms we once again find an odd/even effect. For $\alpha_o$ an odd integer we find
\begin{multline}
 \Tr[e^{i\lambda Q_A}(\tilde\rho^{T_1}_{\rm pure}(t))^{\alpha_o}]=\prod_{k}\prod_{\{x_0|x_t(-k)\in A_1 ~\& ~x_t(k)\in A_2\}}\hspace{-30pt}\Tr[e^{i\lambda(\hat{n}_{x_t}(k)-\hat{n}_{x_t}(-k))}(\rho^{T_1}_{x_0,k}(t))^{\alpha_0}]\\
 =\prod_{k}\prod_{\{x_0|x_t(-k)\in A_1 ~\& ~x_t(k)\in A_2\}}\hspace{-30pt}\big[e^{i\lambda}n(k)^{\alpha_o}+(1-n(k))^{\alpha_o}e^{-i\lambda}\big]
\end{multline}
while for the even case we have 
\begin{align}
 \Tr[e^{i\lambda Q_A}(\tilde\rho^{T_1}_{\rm pure}(t))^{\alpha_e}]
 =\prod_{k}\prod_{\{x_0|x_t(-k)\in A_1 ~\& ~x_t(k)\in A_2\}}\hspace{-30pt}\big[e^{i\lambda}n(k)^{\alpha_e/2}+(1-n(k))^{\alpha_e/2}e^{-i\lambda}\big]^2~.
\end{align}
The same calculation can be performed also using the partial time reversal, whereupon one finds that all the foregoing expressions are the same. Therefore, once again 
\begin{eqnarray}
    N_{\alpha}(t)=N_{\alpha}^{(f)}(t)~~\forall \alpha~.
\end{eqnarray}
Using again the explicit form the counting functions we can proceed as in the case of the R\'enyi negativity~\eqref{eq:renyi_negativity} and recover the results of~\cite{Parez_2022}.

\section{Numerical checks}
\label{sec:numerics}
\begin{figure}[h!]
    \centering
    \includegraphics[width=\linewidth]{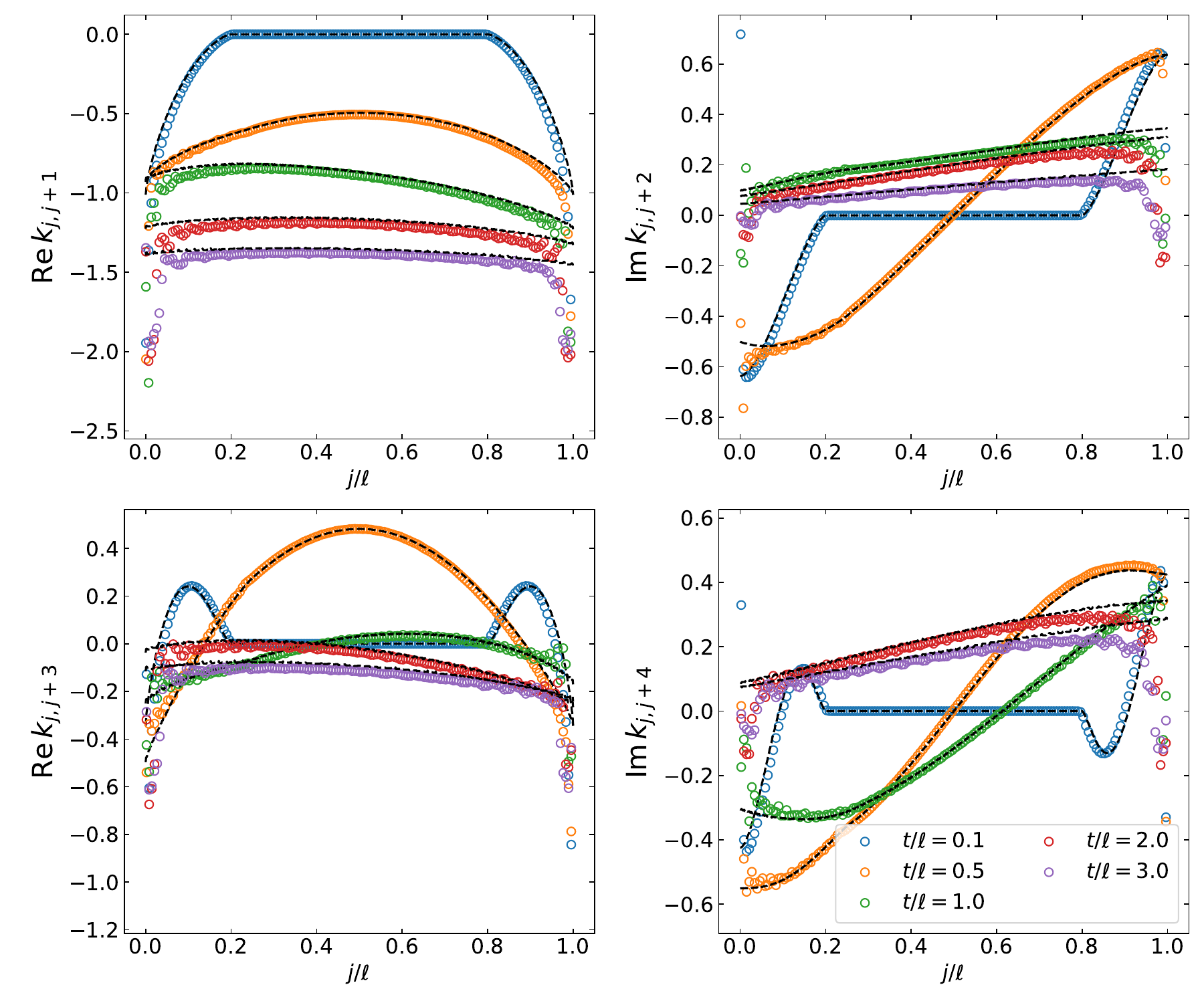}
    \caption{Mixed part of the negativity Hamiltonian (entanglement Hamiltonian) for two intervals of equal lengths $\ell = 800$ placed at a distance $d=600$ (in lattice sites). The symbols are the exact solution \eqref{eq:entanglement_ham} evaluated at different times, while the dashed lines are the quasiparticle predictions \eqref{eq:realspaceeh}. While at small times the result is exactly twice the one for a single interval since there is a maximal speed of propagation $v_{max}=1$, as $t>d$ the presence of the other interval causes an asymmetry which is contained in the counting function of \eqref{eq:realspaceeh}. Here we consider $j\in A_2$.}
    \label{fig:eh}
\end{figure}

Although the analytical checks just performed confirm quite strongly the validity of our approach, in this section we also provide numerical evidence for the structure of negativity and entanglement Hamiltonian which we have obtained. The numerical approach is performed on the same line of \cite{rottoli2024,travaglino2024quasiparticlepictureentanglementHamiltonians}, and exploits the knowledge of the exact time-dependent correlation matrix for the the symmetric state \cite{Eisler_2007}
\begin{equation}
    (C_A)_{x,x'}(t) = \braket{c_x^\dagger c_{x'}} = C_{x,x'}^{(\infty)} + i \frac{x-x'}{4t}e^{-i\frac{\pi}{2}(x+x')} J_{x-x'}(2t)
    \end{equation}
where $C_{x,x'}^{(\infty)} = \frac{1}{2}\delta_{x,x'} + \frac{1}{4}(\delta_{x,x'+1}+\delta_{x,x'-1})$ fixes the long time result, and $J_x(2t)$ is the Bessel function of first kind. The knowledge of the correlation matrix allows to obtain the entanglement Hamiltonian as \cite{Peschel_2009,PhysRevB.69.075111}
\begin{equation}
    C_A(t)  = \frac{1}{1+e^{K_A}} \iff  K_A = \log \frac{1-C_A}{C_A}, \label{eq:entanglement_ham}
\end{equation}
where the equivalence is to be interpreted at the matrix level, considering the coefficients of the entanglement hamiltonian in the basis $c_i^\dagger c_j$, and is clearly not an operator statement.
In our notation, this would correspond to the mixed part of the negativity Hamiltonian. The numerical analysis is shown in figure \ref{fig:eh}, which exhibits perfect agreement between the exact solution and the quasiparticle result. Note that this is not equivalent to the entanglement Hamiltonian which was already in \cite{rottoli2024}, as this refers to the tripartite system. The main difference between the two is the manifest asymmetry of the solution shown in figure \ref{fig:eh}, which is a clear signature of the presence of two intervals $A_1$, $A_2$. 

The numerical evaluation of \eqref{eq:entanglement_ham} is performed through the insertion of a cutoff on the spectrum of the correlation matrix, which as explained in \cite{rottoli2024} has the purpose of removing the contributions of the pure part $\rho_{\rm pure}$. In fact, identifying naturally the pure part as the component which does not give any contribution to the R\'enyi entropies, and recalling that in terms of the eigenvalues $\nu_i$ of the correlation matrix the R\'enyi entropies are expressed as
\begin{equation}
    S^{(\alpha)} = \frac{1}{1-\alpha} \sum_i \log(\nu_i^\alpha+(1-\nu_i)^\alpha)
\end{equation}
we see that the eigenvalues close to 0 and 1 are the ones for which the entanglement contribution disappears, and therefore the cutoff will have to be imposed around these values. This in turn corresponds to eigenvalues of the entanglement Hamiltonian which are very large in modulus, and therefore we see that this is equivalent to a low energy truncation of the entanglement Hamiltonian, as discussed in \cite{rottoli2024}.

In order to obtain the (fermionic) negativity Hamiltonian, the approach is quite similar, upon defining correctly the fermionic transposition of the correlation matrix. For this purpose, it is useful to write the covariance matrix in a form which makes manifest the structure in terms of the blocks $A_1, A_2$,
\begin{equation}
    C_A = \begin{pmatrix}
        C_A^{(1,1)} &C_A^{(1,2)}\\C_A^{(2,1)} & C_A^{(2,2)}   
    \end{pmatrix}.
\end{equation}
In order to allow for the possibility of having terms as $ \braket{c_i^\dagger c^\dagger_j}$ and $ \braket{c_i c_j}$, each component $(C_A)_{ij}$ is to be interpreted as a $2\times2$ matrix containing the four two-point functions:
\begin{equation}
    (C_A)_{ij} = \begin{pmatrix}
        \braket{c_i^\dagger c_j} & \braket{c_i c_j }\\ \braket{c^\dagger_i c^\dagger_j} & \braket{c_i c_j^\dagger}
    \end{pmatrix}.
\end{equation}
From the above discussion, it is clear that the effect of the fermionic transpose on this matrix would be to introduce a modification to all such terms in which $i\in A_1$ and $j\in A_2$ or viceversa, such that 
\begin{equation}
    (C_A)^{R_1}_{i\in A_1,j \in A_2} = \begin{pmatrix}
       i \braket{c_i c_j} & i\braket{c^\dagger_i c_j }\\ i\braket{c_i c^\dagger_j} & i\braket{c^\dagger_i c_j^\dagger}.
    \end{pmatrix}
\end{equation}
In the case of the dimer, since the state is characterized by $\braket{c^\dagger_i c_j^\dagger} =\braket{c_i c_j} =0 $, this would amount to the transformation
\begin{equation}
     (C_A)_{i\in A_1,j \in A_2} =  \begin{pmatrix}
      \braket{c^\dagger_i c_j } & 0\\ 0 & \braket{c_i c_j^\dagger} 
    \end{pmatrix}\to(C_A)^{R_1}_{i\in A_1,j \in A_2} = \begin{pmatrix}
       0 & i\braket{c^\dagger_i c_j }\\ i\braket{c_i c^\dagger_j} & 0
    \end{pmatrix} ,
\end{equation}
which corresponds to 
\begin{equation}
   (C_A)^{R_1}_{i\in A_1,j \in A_2} = i \sigma_x  (C_A)_{i\in A_1,j \in A_2}.
\end{equation}
In terms of this transposed correlation matrix, the negativity Hamiltonian is given  by the immediate generalization of Peschel formula \eqref{eq:entanglement_ham},
\begin{equation}
    \mathcal{N}_{A} = \log \frac{1-C^{R_1}_A}{C^{R_1}_A}. \label{eq:exactN}
\end{equation}
Investigating the structure of the analytical result \eqref{eq:exactN} allows to access the various terms appearing in the negativity Hamiltonian. In particular, the diagonal terms corresponding to the components $c_{x}^\dagger c_{x+z}$ contain both the mixed part $\mathcal{N}_m = K_A(t)$ and the diagonal part $\mathcal{N}_{A,d}(t)$. In order to focus on the latter component, we first remove the imaginary part of the negativity Hamiltonian,
\begin{equation}
    \mathcal{N}_R(t) = \frac{1}{2}\left(\mathcal{N}_A(t) + \mathcal{N}_A^\dagger(t)\right) = \mathcal{N}_{A,d}(t) +  K_A(t)
\end{equation}
and successively remove from this the numerical entanglement Hamiltonian obtained using \eqref{eq:entanglement_ham}, in order to isolate $\mathcal{N}_{A,d}(t)$,
\begin{equation}
    \mathcal{N}_{A,d}(t) =\mathcal{N}_R(t)-  K_A(t)
\end{equation}
\begin{figure}[h!]
    \centering
    \includegraphics[width=0.9\linewidth]{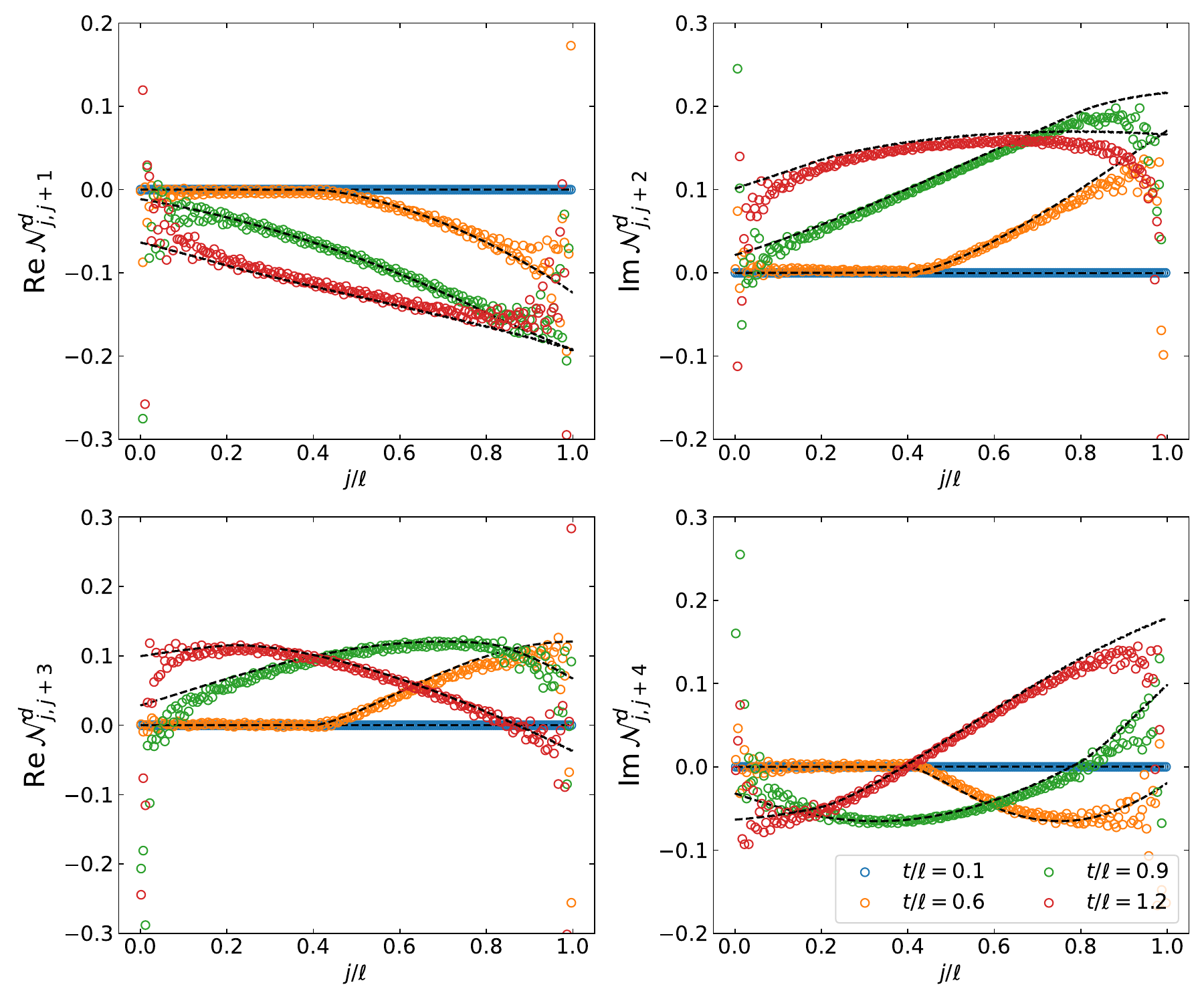}
     \caption{Diagonal part of the negativity Hamiltonian for two intervals of lengths $\ell=1000$ placed at distance $d=600$, with  $j\in A_1$ . The dashed lines represent the analytical prediction \eqref{eq:diagpart}, while the symbols are obtained from the exact solution using \eqref{eq:exactN} and are normalized subtracting $K_A$, which is itself found exactly from \eqref{eq:ent_Ham}.  We see precisely the expected light cone structure arising as the time crosses the value of $d/2$, with correlations emerging only after the quasiparticle pairs are able to couple the two intervals.  }
    \label{fig:diagneg}
\end{figure}
 The result of this analysis is shown in figure \ref{fig:diagneg}, which demonstrates excellent agreement between the numerical evaluation  and the analytical prediction \eqref{eq:diagpart}. In particular, at small times the result is exactly zero, as the pairs have not been able to couple the two intervals yet. At larger times instead a light cone structure emerges as expected from the behaviour of the counting function, since more and more sites in the two intervals become coupled. This light cone effect is completely absent in the entanglement Hamiltonian of figure \ref{fig:eh}, and confirms the interpretation of the two terms as arising from the two distinct classes of pairs, those which are shared between $A_1$ and $A_2$ for $\mathcal{N}_{A,d}(t)$ and those which are shared between $A_1 \cup A_2$ and $\overline{A}$ for $K_A(t)$.

\begin{figure}[h!]
     \centering
     \includegraphics[width=.9\linewidth]{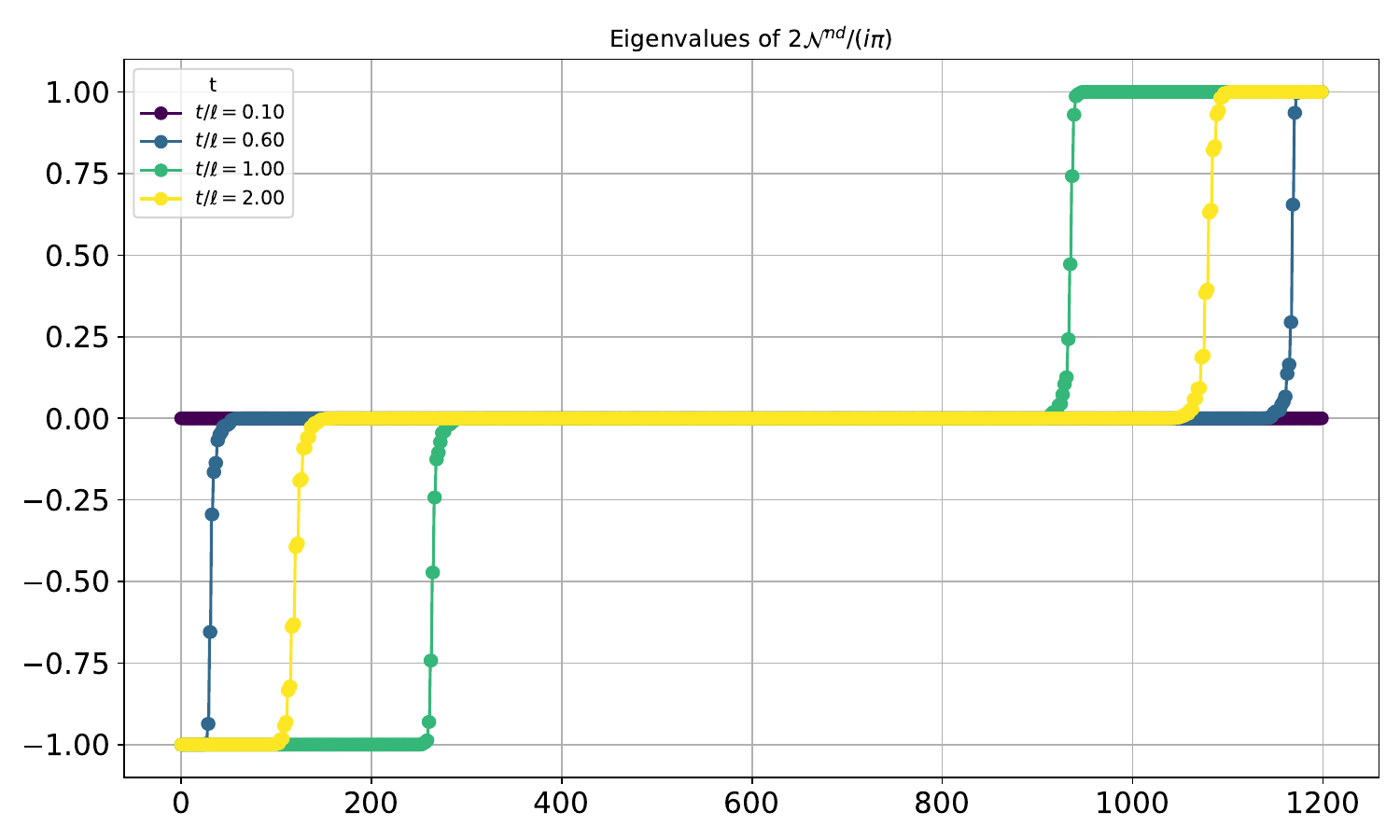}
     \caption{Sorted eigenvalues of the non-diagonal component of the negativity Hamiltonian, normalized by the factor $i \frac{\pi}{2}$, for intervals of equal length $\ell=600$ placed at distance $d=600$. The numerical analysis confirms the theoretical expectation that the eigenvalues of the off-diagonal part are only 0, $\pm \frac{i\pi}{2}$. As expected, at short times the eigenvalues are all zero, while at different times the only variation is provided by the number of nonzero eigenvalues, which initially grows and then decreases back to zero in the infinite time limit.}
     \label{fig:sorted}
 \end{figure}

 The final term which needs to be compared to the numerical results is the non-diagonal part, which has no analogues in any previous quantity considered. This is easily isolated from the rest as it is the imaginary part,
 \begin{equation}
     \mathcal{N}_{A,nd}(t)=i\mathcal{N}_I(t) = \frac{1}{2}\left(\mathcal{N}_A(t) - \mathcal{N}_A^\dagger(t)\right).
 \end{equation}
 As discussed in section \ref{sec:realspace_2}, the main theoretical prediction refers to the single particle eigenvalues of this part of the negativity Hamiltonian, which should only be $0$ and $\pm \frac{i\pi}{2}$, reflecting the algebraic properties of the expression in mixed space representation. Indeed, this is confirmed by the numerical evaluation shown in figure \ref{fig:sorted}. In particular, we see that the only feature distinguishing the result at different times is the number of non-zero eigenvalues: for $t<d/2$ all eigenvalues are zero, as expected from the light cone structure. After a transient growth time, in which this number increases, it starts decreasing again and goes eventually to 0 in the long time limit, as would be expected again from the quasiparticle propagation.
We see therefore that our solution in mixed momentum space/real space representation actually captures the essence of the off-diagonal component, which is given by its eigenvalues, regardless of the fact that the inversion to a full real space description cannot be obtained analytically. Since the eigenvalues (and therefore the phases) are really what determines the structure of the negativities in the hydrodynamic scale, the quasiparticle picture captures the essential physics. It is interesting to note that for small lengths $\ell_1$ and $\ell_2$, the eigenvalues interpolate more smoothly between $-i\pi/2$ and $i\pi/2$, thus leading to a much more complicated structure of possible phases. The main effect of going to the hydrodynamic limit is to restrict to only two phases for the eigenvalues of the partial transposed density matrix, which are $\pm i$.
Since the eigenvalues of the negativity Hamiltonian always come in  complex conjugate pairs \cite{Rottoli2023_finiteT} this implies that when taking traces all imaginary terms would cancel.  That is, they would always appear in pairs of the form 
\begin{equation}
    e^{i\pi/2+ h_k^R} +  e^{-i\pi/2+ h_k^R} = i  e^{ h_k^R} - i  e^{ h_k^R}=0.
\end{equation}
where $h_k^R$ denote the real parts of of the eigenvalues of $\mathcal{N}_A^{(f)}(t)$. This is at the root of the specific form that the even/odd effect of the R\'enyi negativities has in the quasiparticle regime.

\begin{figure}[h!]
    \centering
    \includegraphics[width=\linewidth]{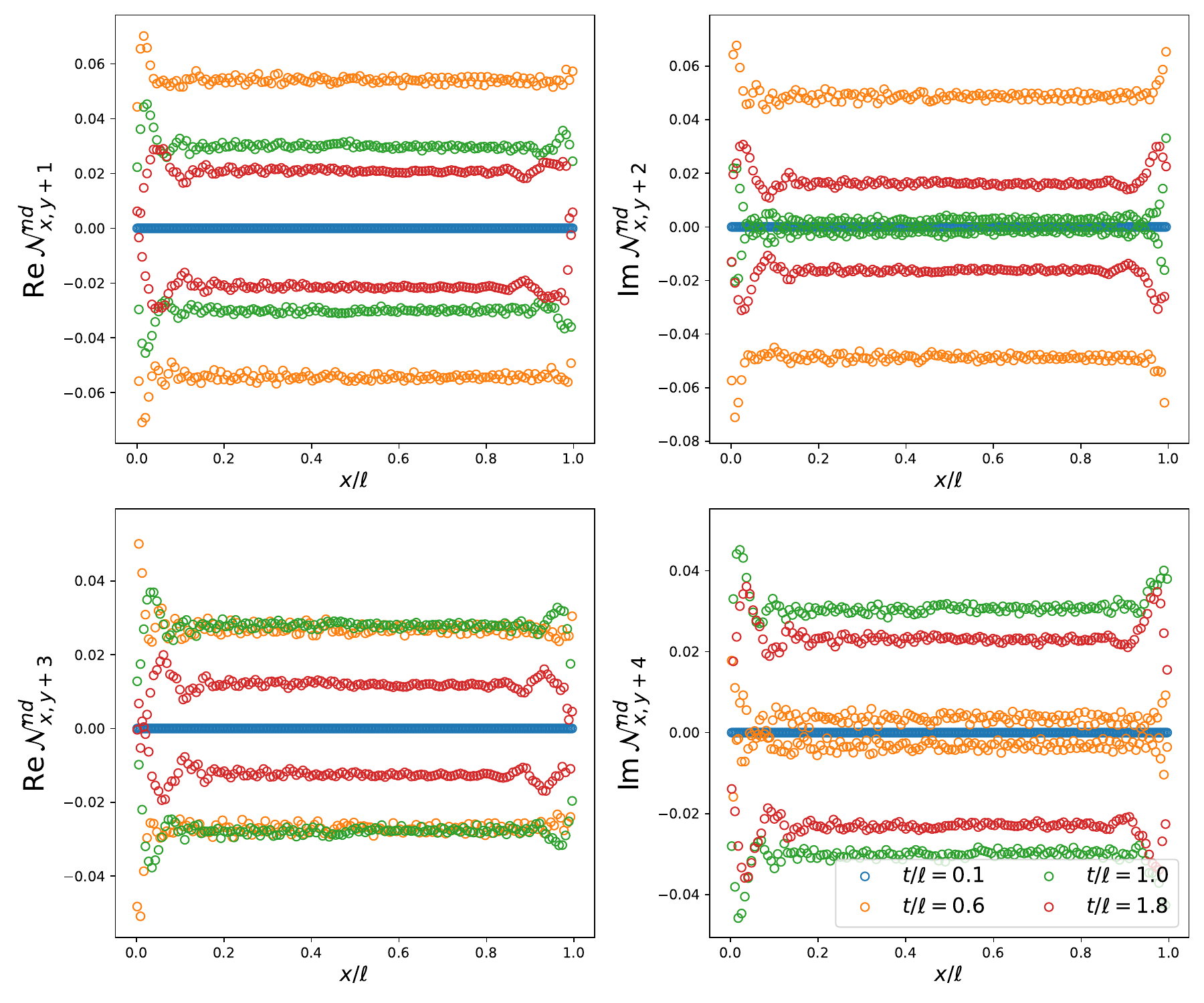}
    \caption{Coefficients of the terms $c_x^\dagger c_{y+z}^\dagger$, where $x\in A_2$ and $y=x-\ell_1-d\in A_1$, for two equal intervals of length $\ell = 800$ placed at distance $d=800$. Except for boundary effects and small oscillations, the solution shows that the coefficients are constant up to an alternating factor $(-1)^x$. Since the chosen poins have fixed distance $x-y = \ell_1 +d +z$, this implies that the kernel is of the form $\mathcal{K}_{nd}(x,y,t) = (-1)^x \mathcal{K}_{nd}(x-y,t)$. }
    \label{fig:offdiag}
\end{figure}

It is still possible and interesting to investigate numerically the precise coefficients of the off-diagonal part, to extrapolate the behaviour of the kernel. Naively, considering the representation of the off-diagonal component in the $b_{x,k}$ representation \eqref{eq:O_dimer_fermion} one would expect to have a coupling which is constant with the distance, since the coupling is between the two modes $b_{x,k}$ and $b_{x-2v_kt,k-\pi}$ without any $x,k$ dependent coefficient. This intuition is almost correct, as shown in figure \ref{fig:offdiag}.
The plot shows clearly that (up to boundary effect and small oscillations) the kernel is of the form 
\begin{equation}
    \mathcal{K}_{nd}(x,y,t) = (-1)^x \mathcal{K}_{nd}(x-y,t)
\end{equation}
namely, an oscillatory factor superimposed to a solution which is constant with the distance. Note that the presence of the $(-1)^x$ factor could also be expected, as it arises naturally from the momentum $k-\pi$ of the left mover. In fact, a similar oscillatory factor is present also in the kernel of the left movers also in the diagonal part of the negativity Hamiltonian, \eqref{eq:kernelminusdimer}. 
Regarding the magnitude of the function $\mathcal{K}_{nd}(x-y,t)$, the main property which has a significant physical interpretation is again its vanishing for $t<d/2$, which is a consequence of the fact that the counting function for this term is the same as for $\mathcal{N}_{A,d}$. At larger times, the magnitude strongly oscillates, but goes back to 0 for $t \to \infty$.

Finally, note that the eigenvalues of the negativity Hamiltonian can be used to obtain the exact solution for the R\'enyi negativities. Considering the definition \eqref{eq:renyinegativitiesdefinition}, and focusing for clarity on the even case, we have 
\begin{equation}
   \mathcal{E}_\alpha= \log \Tr[\left(|\rho^{R_1}|^2\right)^{\alpha/2}] =  \log \Tr[e^{-\alpha \mathcal{N}_R}] - \alpha  \log \Tr[e^{- \mathcal{N}}]
\end{equation}
where $\mathcal{N}_R$ is the real part of the negativity Hamiltonian, and the second term arises from the normalization, since $\rho^{R_1} = e^{-\mathcal{N}}/\Tr[e^{-\mathcal{N}}]$. This can be further simplified upon diagonalization, 
\begin{equation}
     \mathcal{E}_\alpha = \sum_k\left[ \log\left(1+ e^{-\alpha h^{k}_R}\right) - \alpha\log\left(1+ e^{- h^{k}}\right) \right],
\end{equation}
where the sum is over the eigenvalues $h^k$, which are in general complex, and $h^k_R$ is their real part. Note that this is possible because the real and imaginary parts of the negativity Hamiltonian commute, and thus the eigenvalues can be expressed as $h_R + i h_I$ where the real and imaginary parts are eigenvalues of $\mathcal{N}_R$ and $\mathcal{N}_I$ respectively. If $\alpha$ is odd, on the other hand, the definition of negativity implies 
\begin{equation}
  \mathcal{E}_\alpha= \log \Tr[|\rho^{R_1}|^{\alpha-1}\rho^{R_1}] =   \log \Tr[e^{-\alpha \mathcal{N}_R- i \mathcal{N}_I} ] - \alpha  \log \Tr[e^{- \mathcal{N}}],
\end{equation}
which leads to a slightly different result
\begin{equation}
     \mathcal{E}_\alpha = \sum_k\left[ \log\left(1+ e^{-\alpha h^{k}_R - i h_I^k}\right) - \alpha\log\left(1+ e^{- h^{k}}\right) \right].
\end{equation}
This procedure allows one to obtain the exact result which was compared to the quasiparticle prediction in figure \ref{fig:renyinegativities}. This provides a further and final demonstration of the validity of the negativity Hamiltonian.

\section{Conclusions}
\label{sec:conclusions}
In this paper, we have extended the framework introduced in Ref. \cite{rottoli2024}—originally developed for the study of the entanglement Hamiltonian—to systems involving more complex spatial partitions. This generalization has enabled us to investigate the structure of the negativity Hamiltonian, which governs entanglement in mixed states. In particular, we have used an emergent hydrodynamical picture to characterize the negativity and entanglement Hamiltonians in tripartite geometries, obtaining in a simple and concise way quasiparticle predictions for the relevant entanglement measures, which were in some cases obtained in previous works. This framework offers a natural interpretation of certain important, yet not fully understood, observations—such as the equivalence between standard Rényi negativities and fermionic negativities at the ballistic scale. 
Hence, this work confirms that the hydrodynamic approach developed in Ref. \cite{rottoli2024} offers the appropriate framework within which the quasiparticle picture should be understood and interpreted.

This approach—and in particular, the simple tensor structure of the density matrix in Eq.~\eqref{eq_decomposition}—provides an ideal framework for investigating the out-of-equilibrium behavior of various physically relevant quantities at the hydrodynamic scale. Natural candidates for such analysis include the full counting statistics of conserved charges and currents, both in homogeneous and inhomogeneous settings. While the inhomogeneous case requires additional care to account for the space-time dependence of the occupation functions, the core principles of our approach can be extended to these scenarios with only minor modifications \cite{Bertini_2018}.

Some further interesting lines of investigation include the generalization of the hydrodynamic framework to dissipative systems, for which a suitable modification of the quasiparticle picture can be introduced in some situations\cite{alba_carollo2021, carollo_alba2022, Alba_2022}. 
Another generalization concerns monitored systems, in which it is yet unclear whether the interplay between measurements and unitary evolution can be recast in terms of some emergent theory \cite{carollo2022}. 
In particular, a formulation in terms of our hydrodynamic density matrix could shed light on the problem and suggest a possible quasiparticle solution which has remained elusive until now.
Finally, extending this discussion to interacting systems—at least in the case of integrable models in (1+1) dimensions—could provide valuable insights into the limitations of the quasiparticle picture, whose breakdown has been observed in certain contexts \cite{PRX}. Such an investigation may also clarify the precise relationship between the quasiparticle framework and the emerging spacetime duality approach, potentially bridging the gap between these two perspectives on entanglement dynamics.

\medskip

 \noindent {\bf Acknowledgments:} 
We thank for useful discussions Filiberto Ares, Angelo Russotto, Andrea Stampiggi, Federico Rottoli and Viktor Eisler.
PC and CR acknowledge support from European Union-NextGenerationEU, in the framework of the PRIN 2022 Project HIGHEST no. 2022SJCKAH\_002.

\appendix
\section{Quasiparticle counting}
In this section we perform explicitly the quasiparticle counting, which allows to fix the prefactors $\chi_{\rm p}(x,k,t)$ and $\chi_{\rm m}(x,k,t)$.
\label{appA}
\subsection{Pure part}
In this case, we are interested in the pairs in which the right mover is in $A_2$ and the left mover is in $A_1$. Considering for simplicity two intervals of equal length $\ell$ at distance d (the extension to arbitrary lengths is immediate), and placing the origin in the middle, the condition reduces to 
\begin{equation}
    \begin{cases}
        x_+ \in [d/2,d/2+\ell]\\
        x_- \in [-d/2-\ell,-d/2]
    \end{cases}\Rightarrow 
        x_+ \in [d/2,d/2+\ell] \cap [-d/2-\ell + 2v_kt,-d/2+2v_kt] \nonumber
\end{equation}
Where we take $v_k$ as the positive velocity. Note that the intersection is zero unless $\frac{d}{2}< v_kt <\ell+\frac{d}{2}$, and this implies that at long times the pure contribution disappears. If the contribution is nonzero then the intersection is given by
\begin{equation}
    x \in [\max(d/2,-\ell-d/2+2v_kt), \min(d/2+\ell,-d/2-\ell +2v_kt)] .
\end{equation}
This immediately fixes the bounds of the product (or of the integral if we consider the negativity Hamiltonian),
\begin{equation}
    \rho_{pure}^{T_1} = \prod_{k>0} \prod_{x = \max(d/2,-\ell-d/2+2vt)}^{\min(d/2+\ell,-d/2 +2vt)} \rho_{x,k}^{(t)} \chi_{[\frac{d}{2},\ell+\frac{d}{2}]}(v_kt).
\end{equation}
When taking traces, this gives the expected result
\beqa
    R^{(n)} = \int_{k>0}\frac{dk}{2\pi} \left[\min(\frac{d}{2}+\ell,-\frac{d}{2} +2v_kt) - \max(\frac{d}{2},-\ell-\frac{d}{2}+2v_kt)\right]  \chi_{[\frac{d}{2},\ell+\frac{d}{2}]}(v_kt) \tilde{s}_\alpha(k) ],
    \label{eq:r}
\eeqa
where $\tilde{s}_\alpha(k)$ was introduced in \eqref{eq:varepsilon}.
 This is exactly the same result as in \cite{Murciano2022}, although this is not immediate to see, since
\beqa
    \left[\min(d/2+\ell,-d/2 +2vt) - \max(d/2,-\ell-d/2+2vt)\right] \chi_{[\frac{d}{2},\ell+\frac{d}{2}]}(v_kt)\nonumber\\ = 2 \left[\max(vt,d/2)-2\max(vt,(l+d)/2) + \max(vt,l+d/2)\right] ,\nonumber
\eeqa
where the factor 2 is present since in \eqref{eq:r} the integration is restricted to positive momenta. Note that the result can be written without the introduction of the characteristic function using as extremes of integration 
 \begin{equation}
    \rho_{pure}^{T_1} = \prod_{k>0} \hspace{0.3cm}\prod_{x = \min(\max(\frac{d}{2},-\ell-\frac{d}{2}+2vt),\frac{d}{2}+\ell)}^{\max(\min(\frac{d}{2}+\ell,-\frac{d}{2} +2vt),\frac{d}{2})} \rho_{x,k}(t) .
\end{equation}
It can be checked that this is the same as the above.
\subsection{Mixed part}
For the mixed part, the counting functions selects two contributions: 
\begin{equation}
    (x_\pm \in A_1 \cap x_\mp \in \overline{A}) \cup  (x_\pm \in A_2 \cap x_\mp \in \overline{A})
    \label{eq:set}
\end{equation}
We are interested in the evaluation of
\begin{equation}
    \int dx \chi_{\rm m}(x,k,t)
\end{equation}
which is needed to evaluate the R\'enyi entropies as
\begin{equation}
    S_A^\alpha(t) =  \int \frac{dk}{2\pi} \left(\int dx \chi_{\rm m}(x,k,t)\right) \log(n(k)^{\alpha} + (1-n(k))^{\alpha}) 
    \label{eq:entropy}.
\end{equation}
Which consequently gives also the R\'enyi negativities from \eqref{eq:renyi_negativity}. The structure of \eqref{eq:set} can be expressed as
\begin{equation}
    (x_\pm \in A_1 \cap x_\mp \in \overline{A}) = (x_\pm \in A_1 \cap x_\mp \in \overline{A}_1) -(x_\pm \in A_1 \cap x_\mp \in A_2),  
\end{equation}
namely, we just need to subtract the terms which are shared between the two subcomponents from the full contributions of terms which are shared between, for instance, $A_1$ and its full complement. But for these two quantities the counting is immediate: the full number of pairs which are shared between $A_1$ and $\overline{A}_1$ is the usual term which appears in standard quasiparticle picture applications:
\begin{equation}
    \min(2|v_k|t,\ell) = 2|v_k|t +\ell - \max(2|v_k|t,\ell)
\end{equation}
where the right hand side has been introduced juts to match the notation of \cite{Murciano2022}. Regarding the second part, it is simply what was discussed above for the pure part. Subtracting the two contributions, and observing that the two intervals contribute the same factor, we get precisely 
\beqa
    \int dx \chi_{\rm m}(x,k,t)&=& 2 \left( 2|v_k|t +\ell - \max(2|v_k|t,\ell)\right) \hspace{5cm}\\&-& 2\left[\max(|v_k|t,d/2)-2\max(|v_k|t,(l+d)/2) + \max(|v_k|t,l+d/2)\right] \nonumber
\eeqa
which reproduces exactly the result of \cite{Murciano2022}. 

In the long lime limit, $t \gg \ell$, this saturates to $2\ell$. This reflects the fact that the system saturates to two separate GGEs, one for each interval, leading precisely to the form \eqref{eq:gge}.

\bibliographystyle{ytphys}
\bibliography{bibliography}

\providecommand{\href}[2]{#2}\begingroup\begin{thebibliography}{10}

\bibitem{vidal2003}
G.~Vidal, J.~I. Latorre, E.~Rico, and A.~Kitaev, {\slshape Entanglement in quantum critical phenomena,} \href{https://link.aps.org/doi/10.1103/PhysRevLett.90.227902}{{\em Phys. Rev. Lett.} {\bfseries 90} (2003) 227902}.

\bibitem{Plenio2005AnIT}
M.~B. Plenio and S.~Virmani, {\slshape An introduction to entanglement measures,} \href{https://api.semanticscholar.org/CorpusID:7131013}{{\em Quantum Inf. Comput.} {\bfseries 7} (2005) 1--51}.

\bibitem{damico2008}
L.~Amico, R.~Fazio, A.~Osterloh, and V.~Vedral, {\slshape Entanglement in many-body systems,} \href{https://link.aps.org/doi/10.1103/RevModPhys.80.517}{{\em Rev. Mod. Phys.} {\bfseries 80} (2008) 517--576}.

\bibitem{horodecki2009}
R.~Horodecki, P.~Horodecki, M.~Horodecki, and K.~Horodecki, {\slshape Quantum entanglement,} \href{https://link.aps.org/doi/10.1103/RevModPhys.81.865}{{\em Rev. Mod. Phys.} {\bfseries 81} (2009) 865--942}.

\bibitem{Calabrese_2009_1}
P.~Calabrese, J.~Cardy, and B.~Doyon, {\slshape Entanglement entropy in extended quantum systems,} \href{https://dx.doi.org/10.1088/1751-8121/42/50/500301}{{\em J. Phys. A Math. Theor.} {\bfseries 42} (2009) 500301}.

\bibitem{arealaws}
J.~Eisert, M.~Cramer, and M.~B. Plenio, {\slshape Colloquium: Area laws for the entanglement entropy,} \href{https://link.aps.org/doi/10.1103/RevModPhys.82.277}{{\em Rev. Mod. Phys.} {\bfseries 82} (2010) 277--306}.

\bibitem{Pasquale_Calabrese_2004}
P.~Calabrese and J.~Cardy, {\slshape Entanglement entropy and quantum field theory,} \href{https://dx.doi.org/10.1088/1742-5468/2004/06/P06002}{{\em J. Stat. Mech.} (2004) P06002}.

\bibitem{Calabrese_2009}
P.~Calabrese and J.~Cardy, {\slshape Entanglement entropy and conformal field theory,} \href{https://dx.doi.org/10.1088/1751-8113/42/50/504005}{{\em J. Phys. A Math. Theor.} {\bfseries 42} (2009) 504005}.

\bibitem{Deutsch1991}
J.~M. Deutsch, {\slshape Quantum statistical mechanics in a closed system,} \href{https://link.aps.org/doi/10.1103/PhysRevA.43.2046}{{\em Phys. Rev. A} {\bfseries 43} (1991) 2046}.

\bibitem{srednicki1}
M.~Srednicki, {\slshape Chaos and quantum thermalization,} \href{https://link.aps.org/doi/10.1103/PhysRevE.50.888}{{\em Phys. Rev. E} {\bfseries 50} (1994) 888}.

\bibitem{Rigol:2007juv}
M.~Rigol, V.~Dunjko, and M.~Olshanii, {\slshape {Thermalization and its mechanism for generic isolated quantum systems},} \href{http://dx.doi.org/10.1038/nature06838}{{\em Nature} {\bfseries 452} (2008) 854}.

\bibitem{polkovnikov}
A.~Polkovnikov, K.~Sengupta, A.~Silva, and M.~Vengalattore, {\slshape Colloquium: Nonequilibrium dynamics of closed interacting quantum systems,} \href{https://link.aps.org/doi/10.1103/RevModPhys.83.863}{{\em Rev. Mod. Phys.} {\bfseries 83} (2011) 863}.

\bibitem{DAlessio:2015qtq}
L.~D'Alessio, Y.~Kafri, A.~Polkovnikov, and M.~Rigol, {\slshape {From quantum chaos and eigenstate thermalization to statistical mechanics and thermodynamics},} \href{http://dx.doi.org/10.1080/00018732.2016.1198134}{{\em Adv. Phys.} {\bfseries 65} (2016) 239}.

\bibitem{Gogolin_2016}
C.~Gogolin and J.~Eisert, {\slshape Equilibration, thermalisation, and the emergence of statistical mechanics in closed quantum systems,} \href{https://dx.doi.org/10.1088/0034-4885/79/5/056001}{{\em Rep. Prog. Phys.} {\bfseries 79} (2016) 056001}.

\bibitem{Essler_quench}
F.~H.~L. Essler and M.~Fagotti, {\slshape {Quench Dynamics and Relaxation in Isolated Integrable Quantum Spin Chains},} \href{https://dx.doi.org/10.1088/1742-5468/2016/06/064002}{{\em J. Stat. Mech.} (2016) 064002}.

\bibitem{Calabrese_2016}
P.~Calabrese, F.~H.~L. Essler, and G.~Mussardo, {\slshape Introduction to quantum integrability in out of equilibrium systems,} \href{https://dx.doi.org/10.1088/1742-5468/2016/06/064001}{{\em J. Stat. Mech.} (2016) 064001}.

\bibitem{Collura_2014}
M.~Collura, M.~Kormos, and P.~Calabrese, {\slshape Stationary entanglement entropies following an interaction quench in 1d bose gas,} \href{https://dx.doi.org/10.1088/1742-5468/2014/01/P01009}{{\em J. Stat. Mech.} (2014) P01009}.

\bibitem{quench2}
P.~Calabrese and J.~Cardy, {\slshape Evolution of entanglement entropy in one-dimensional systems,} \href{https://dx.doi.org/10.1088/1742-5468/2005/04/P04010}{{\em J. Stat. Mech.} (2005) P04010}.

\bibitem{Bisognano:1975ih}
J.~J. Bisognano and E.~H. Wichmann, {\slshape {On the Duality Condition for a Hermitian Scalar Field},} \href{http://dx.doi.org/10.1063/1.522605}{{\em J. Math. Phys.} {\bfseries 16} (1975) 985}.

\bibitem{Bisognano:1976za}
J.~J. Bisognano and E.~H. Wichmann, {\slshape {On the Duality Condition for Quantum Fields},} \href{http://dx.doi.org/10.1063/1.522898}{{\em J. Math. Phys.} {\bfseries 17} (1976) 303}.

\bibitem{Cardy_2016}
J.~Cardy and E.~Tonni, {\slshape Entanglement hamiltonians in two-dimensional conformal field theory,} \href{https://dx.doi.org/10.1088/1742-5468/2016/12/123103}{{\em J. Stat. Mech.} (2016) 123103}.

\bibitem{VidalPRA2002}
G.~Vidal and R.~F. Werner, {\slshape Computable measure of entanglement,} \href{https://link.aps.org/doi/10.1103/PhysRevA.65.032314}{{\em Phys. Rev. A} {\bfseries 65} (2002) 032314}.

\bibitem{Plenio2005}
M.~B. Plenio, {\slshape Logarithmic negativity: A full entanglement monotone that is not convex,} \href{https://link.aps.org/doi/10.1103/PhysRevLett.95.090503}{{\em Phys. Rev. Lett.} {\bfseries 95} (2005) 090503}.

\bibitem{PeresPRL1996}
A.~Peres, {\slshape Separability criterion for density matrices,} \href{https://link.aps.org/doi/10.1103/PhysRevLett.77.1413}{{\em Phys. Rev. Lett.} {\bfseries 77} (1996) 1413--1415}.

\bibitem{Simon2000}
R.~Simon, {\slshape Peres-horodecki separability criterion for continuous variable systems,} \href{https://link.aps.org/doi/10.1103/PhysRevLett.84.2726}{{\em Phys. Rev. Lett.} {\bfseries 84} (2000) 2726--2729}.

\bibitem{calabrese_cardy_tonni_2012}
P.~Calabrese, J.~Cardy, and E.~Tonni, {\slshape Entanglement negativity in quantum field theory,} \href{https://link.aps.org/doi/10.1103/PhysRevLett.109.130502}{{\em Phys. Rev. Lett.} {\bfseries 109} (2012) 130502}.

\bibitem{Calabrese_cardy_tonni_2013}
P.~Calabrese, J.~Cardy, and E.~Tonni, {\slshape Entanglement negativity in extended systems: a field theoretical approach,} \href{https://dx.doi.org/10.1088/1742-5468/2013/02/P02008}{{\em J. Stat. Mech.} (2013) P02008}.

\bibitem{Capizzi_2022}
L.~Capizzi, S.~Murciano, and P.~Calabrese, {\slshape Rényi entropy and negativity for massless complex boson at conformal interfaces and junctions,} \href{http://dx.doi.org/10.1007/JHEP11(2022)105}{{\em J. High Energy Phys.} {\bfseries 11} (2022) 105}.

\bibitem{Capizzi_2022b}
L.~Capizzi, S.~Murciano, and P.~Calabrese, {\slshape Rényi entropy and negativity for massless dirac fermions at conformal interfaces and junctions,} \href{http://dx.doi.org/10.1007/JHEP08(2022)171}{{\em J. High Energy Phys.} {\bfseries 08} (2022) 171}.

\bibitem{Calabrese_2014}
P.~Calabrese, J.~Cardy, and E.~Tonni, {\slshape Finite temperature entanglement negativity in conformal field theory,} \href{http://dx.doi.org/10.1088/1751-8113/48/1/015006}{{\em J. Phys. A} {\bfseries 48} (2014) 015006}.

\bibitem{Blondeau_2016}
O.~Blondeau-Fournier, O.~A. Castro-Alvaredo, and B.~Doyon, {\slshape Universal scaling of the logarithmic negativity in massive quantum field theory,} \href{http://dx.doi.org/10.1088/1751-8113/49/12/125401}{{\em J. Phys. A} {\bfseries 49} (2016) 125401}.

\bibitem{Castro_Alvaredo_2019}
O.~A. Castro-Alvaredo, C.~De~Fazio, B.~Doyon, and I.~M. Szécsényi, {\slshape Entanglement content of quantum particle excitations. part ii. disconnected regions and logarithmic negativity,} \href{http://dx.doi.org/10.1007/JHEP11(2019)058}{{\em J. High Energy Phys.} {\bfseries 11} (2019) 58}.

\bibitem{neg_spec}
P.~Ruggiero, V.~Alba, and P.~Calabrese, {\slshape Negativity spectrum of one-dimensional conformal field theories,} \href{https://link.aps.org/doi/10.1103/PhysRevB.94.195121}{{\em Phys. Rev. B} {\bfseries 94} (2016) 195121}.

\bibitem{negativity_harmonic}
K.~Audenaert, J.~Eisert, M.~B. Plenio, and R.~F. Werner, {\slshape Entanglement properties of the harmonic chain,} \href{https://link.aps.org/doi/10.1103/PhysRevA.66.042327}{{\em Phys. Rev. A} {\bfseries 66} (2002) 042327}.

\bibitem{Cavalcanti_2008}
D.~Cavalcanti, A.~Ferraro, A.~Garc\'{\i}a-Saez, and A.~Ac\'{\i}n, {\slshape Distillable entanglement and area laws in spin and harmonic-oscillator systems,} \href{https://link.aps.org/doi/10.1103/PhysRevA.78.012335}{{\em Phys. Rev. A} {\bfseries 78} (2008) 012335}.

\bibitem{Eisler_2014}
V.~Eisler and Z.~Zimborás, {\slshape Entanglement negativity in the harmonic chain out of equilibrium,} \href{https://dx.doi.org/10.1088/1367-2630/16/12/123020}{{\em New J. Phys.} {\bfseries 16} (2014) 123020}.

\bibitem{negativity_random}
P.~Ruggiero, V.~Alba, and P.~Calabrese, {\slshape Entanglement negativity in random spin chains,} \href{https://link.aps.org/doi/10.1103/PhysRevB.94.035152}{{\em Phys. Rev. B} {\bfseries 94} (2016) 035152}.

\bibitem{spin3}
X.~Turkeshi, P.~Ruggiero, and P.~Calabrese, {\slshape Negativity spectrum in the random singlet phase,} \href{https://link.aps.org/doi/10.1103/PhysRevB.101.064207}{{\em Phys. Rev. B} {\bfseries 101} (2020) 064207}.

\bibitem{Wybo_2021}
E.~Wybo, M.~Knap, and F.~Pollmann, {\slshape Dynamics of negativity of a wannier–stark many‐body localized system coupled to a bath,} \href{http://dx.doi.org/10.1002/pssb.202100161}{{\em Phys. Status Solidi B} {\bfseries 259} (2021) }.

\bibitem{spin1}
R.~A. Santos, V.~Korepin, and S.~Bose, {\slshape Negativity for two blocks in the one-dimensional spin-1 affleck-kennedy-lieb-tasaki model,} \href{https://link.aps.org/doi/10.1103/PhysRevA.84.062307}{{\em Phys. Rev. A} {\bfseries 84} (2011) 062307}.

\bibitem{spin2}
T.-C. Lu and T.~Grover, {\slshape Singularity in entanglement negativity across finite-temperature phase transitions,} \href{https://link.aps.org/doi/10.1103/PhysRevB.99.075157}{{\em Phys. Rev. B} {\bfseries 99} (2019) 075157}.

\bibitem{spin4}
G.~B. Mbeng, V.~Alba, and P.~Calabrese, {\slshape Negativity spectrum in 1d gapped phases of matter,} \href{https://dx.doi.org/10.1088/1751-8121/aa6734}{{\em J. Phys. A Math. Theor.} {\bfseries 50} (2017) 194001}.

\bibitem{Rogerson_2022}
D.~Rogerson, F.~Pollmann, and A.~Roy, {\slshape Entanglement entropy and negativity in the ising model with defects,} \href{http://dx.doi.org/10.1007/JHEP06(2022)165}{{\em J. High Energy Phys.} {\bfseries 06} (2022) 165}.

\bibitem{Choi_2024}
W.~Choi, M.~Knap, and F.~Pollmann, {\slshape Finite-temperature entanglement negativity of fermionic symmetry-protected topological phases and quantum critical points in one dimension,} \href{http://dx.doi.org/10.1103/PhysRevB.109.115132}{{\em Phys. Rev. B} {\bfseries 109} (2024) }.

\bibitem{Elben_2020}
A.~Elben, R.~Kueng, H.-Y.~R. Huang, R.~van Bijnen, C.~Kokail, M.~Dalmonte, P.~Calabrese, B.~Kraus, J.~Preskill, P.~Zoller, and B.~Vermersch, {\slshape Mixed-state entanglement from local randomized measurements,} \href{https://link.aps.org/doi/10.1103/PhysRevLett.125.200501}{{\em Phys. Rev. Lett.} {\bfseries 125} (2020) 200501}.

\bibitem{Neven_2021}
A.~Neven, J.~Carrasco, V.~Vitale, C.~Kokail, A.~Elben, M.~Dalmonte, P.~Calabrese, P.~Zoller, B.~Vermersch, R.~Kueng, and B.~Kraus, {\slshape Symmetry-resolved entanglement detection using partial transpose moments,} \href{http://dx.doi.org/10.1038/s41534-021-00487-y}{{\em npj Quantum Inf.} {\bfseries 7} (2021) }.

\bibitem{Elben_2022}
A.~Elben, S.~T. Flammia, H.-Y. Huang, R.~Kueng, J.~Preskill, B.~Vermersch, and P.~Zoller, {\slshape The randomized measurement toolbox,} \href{http://dx.doi.org/10.1038/s42254-022-00535-2}{{\em Nat. Rev. Phys.} {\bfseries 5} (2022) 9–24}.

\bibitem{Calabrese_tagliacozzo_tonni_2013}
P.~Calabrese, L.~Tagliacozzo, and E.~Tonni, {\slshape Entanglement negativity in the critical ising chain,} \href{https://dx.doi.org/10.1088/1742-5468/2013/05/P05002}{{\em J. Stat. Mech.} (2013) P05002}.

\bibitem{Alba_2013}
V.~Alba, {\slshape Entanglement negativity and conformal field theory: a monte carlo study,} \href{https://dx.doi.org/10.1088/1742-5468/2013/05/P05013}{{\em J. Stat. Mech.} {\bfseries 2013} (2013) P05013}.

\bibitem{chung_alba_2014}
C.-M. Chung, V.~Alba, L.~Bonnes, P.~Chen, and A.~M. L\"auchli, {\slshape Entanglement negativity via the replica trick: A quantum monte carlo approach,} \href{https://link.aps.org/doi/10.1103/PhysRevB.90.064401}{{\em Phys. Rev. B} {\bfseries 90} (2014) 064401}.

\bibitem{shapourian2017many}
H.~Shapourian, K.~Shiozaki, and S.~Ryu, {\slshape Many-body topological invariants for fermionic symmetry-protected topological phases,} \href{http://dx.doi.org/10.1103/PhysRevLett.118.216402}{{\em Phys. Rev. Lett.} {\bfseries 118} (2017) }.

\bibitem{eisler2016negativity2d}
V.~Eisler and Z.~Zimbor\'as, {\slshape Entanglement negativity in two-dimensional free lattice models,} \href{https://link.aps.org/doi/10.1103/PhysRevB.93.115148}{{\em Phys. Rev. B} {\bfseries 93} (2016) 115148}.

\bibitem{Shapourian2017}
H.~Shapourian, K.~Shiozaki, and S.~Ryu, {\slshape Partial time-reversal transformation and entanglement negativity in fermionic systems,} \href{https://link.aps.org/doi/10.1103/PhysRevB.95.165101}{{\em Phys. Rev. B} {\bfseries 95} (2017) 165101}.

\bibitem{Shapourian_2019}
H.~Shapourian and S.~Ryu, {\slshape Finite-temperature entanglement negativity of free fermions,} \href{https://dx.doi.org/10.1088/1742-5468/ab11e0}{{\em J. Stat. Mech.} (2019) 043106}.

\bibitem{shapourian_ruggiero}
H.~Shapourian, P.~Ruggiero, S.~Ryu, and P.~Calabrese, {\slshape {Twisted and untwisted negativity spectrum of free fermions},} \href{https://scipost.org/10.21468/SciPostPhys.7.3.037}{{\em SciPost Phys.} {\bfseries 7} (2019) 037}.

\bibitem{Cornfeld2019}
E.~Cornfeld, E.~Sela, and M.~Goldstein, {\slshape Measuring fermionic entanglement: Entropy, negativity, and spin structure,} \href{https://link.aps.org/doi/10.1103/PhysRevA.99.062309}{{\em Phys. Rev. A} {\bfseries 99} (2019) 062309}.

\bibitem{shapourian2018negativity}
H.~Shapourian and S.~Ryu, {\slshape Entanglement negativity of fermions: Monotonicity, separability criterion, and classification of few-mode states,} \href{https://link.aps.org/doi/10.1103/PhysRevA.99.022310}{{\em Phys. Rev. A} {\bfseries 99} (2019) 022310}.

\bibitem{murcianonegativityHamiltonian}
S.~Murciano, V.~Vitale, M.~Dalmonte, and P.~Calabrese, {\slshape Negativity hamiltonian: An operator characterization of mixed-state entanglement,} \href{https://link.aps.org/doi/10.1103/PhysRevLett.128.140502}{{\em Phys. Rev. Lett.} {\bfseries 128} (2022) 140502}.

\bibitem{Rottoli_2023}
F.~Rottoli, S.~Murciano, E.~Tonni, and P.~Calabrese, {\slshape Entanglement and negativity hamiltonians for the massless dirac field on the half line,} \href{https://dx.doi.org/10.1088/1742-5468/acb262}{{\em J. Stat. Mech.} (2023) 013103}.

\bibitem{Rottoli2023_finiteT}
F.~Rottoli, S.~Murciano, and P.~Calabrese, {\slshape Finite temperature negativity hamiltonians of the massless dirac fermion,} \href{https://doi.org/10.1007/JHEP06(2023)139}{{\em J. High Energy Phys.} {\bfseries 06} (2023) 139}.

\bibitem{bertini2018entanglement}
B.~Bertini, E.~Tartaglia, and P.~Calabrese, {\slshape Entanglement and diagonal entropies after a quench with no pair structure,} \href{http://dx.doi.org/10.1088/1742-5468/aac73f}{{\em J. Stat. Mech.} (2018) 063104}.

\bibitem{bastianello_pair}
A.~Bastianello and M.~Collura, {\slshape {Entanglement spreading and quasiparticle picture beyond the pair structure},} \href{https://scipost.org/10.21468/SciPostPhys.8.3.045}{{\em SciPost Phys.} {\bfseries 8} (2020) 045}.

\bibitem{alba1}
V.~Alba and P.~Calabrese, {\slshape Entanglement and thermodynamics after a quantum quench in integrable systems,} \href{http://dx.doi.org/10.1073/pnas.1703516114}{{\em PNAS} {\bfseries 114} (2017) 7947}.

\bibitem{alba2}
V.~Alba and P.~Calabrese, {\slshape {Entanglement dynamics after quantum quenches in generic integrable systems},} \href{https://scipost.org/10.21468/SciPostPhys.4.3.017}{{\em SciPost Phys.} {\bfseries 4} (2018) 017}.

\bibitem{Alba_2019}
V.~Alba and P.~Calabrese, {\slshape Quantum information dynamics in multipartite integrable systems,} \href{https://dx.doi.org/10.1209/0295-5075/126/60001}{{\em Europhys. Lett.} {\bfseries 126} (2019) 60001}.

\bibitem{parez2021quasiparticle}
G.~Parez, R.~Bonsignori, and P.~Calabrese, {\slshape Quasiparticle dynamics of symmetry-resolved entanglement after a quench: Examples of conformal field theories and free fermions,} \href{https://link.aps.org/doi/10.1103/PhysRevB.103.L041104}{{\em Phys. Rev. B} {\bfseries 103} (2021) L041104}.

\bibitem{Coser_2014}
A.~Coser, E.~Tonni, and P.~Calabrese, {\slshape Entanglement negativity after a global quantum quench,} \href{http://dx.doi.org/10.1088/1742-5468/2014/12/P12017}{{\em J. Stat. Mech.} (2014) P12017}.

\bibitem{Murciano2022}
S.~Murciano, V.~Alba, and P.~Calabrese, {\em Quench Dynamics of R{\'e}nyi Negativities and the Quasiparticle Picture}, \href{http://dx.doi.org/10.1007/978-3-031-03998-0_14}{pp.~397--424}.
\newblock Springer International Publishing, 2022.

\bibitem{Parez_2022}
G.~Parez, R.~Bonsignori, and P.~Calabrese, {\slshape Dynamics of charge-imbalance-resolved entanglement negativity after a quench in a free-fermion model,} \href{https://dx.doi.org/10.1088/1742-5468/ac666c}{{\em J. Stat. Mech.} (2022) 053103}.

\bibitem{groha2018full}
S.~Groha, F.~Essler, and P.~Calabrese, {\slshape {Full counting statistics in the transverse field Ising chain},} \href{http://dx.doi.org/10.21468/SciPostPhys.4.6.043}{{\em SciPost Phys.} {\bfseries 4} (2018) 043}.

\bibitem{horvath2024full}
D.~X. Horv\'ath and C.~Rylands, {\slshape Full counting statistics of charge in quenched quantum gases,} \href{https://link.aps.org/doi/10.1103/PhysRevA.109.043302}{{\em Phys. Rev. A} {\bfseries 109} (2024) 043302}.

\bibitem{rottoli2024}
F.~Rottoli, C.~Rylands, and P.~Calabrese, {\slshape Entanglement hamiltonians and the quasiparticle picture,} \href{https://link.aps.org/doi/10.1103/PhysRevB.111.L140302}{{\em Phys. Rev. B} {\bfseries 111} (2025) L140302}.

\bibitem{travaglino2024quasiparticlepictureentanglementHamiltonians}
R.~Travaglino, C.~Rylands, and P.~Calabrese, {\slshape Quasiparticle picture for entanglement hamiltonians in higher dimensions,} \href{https://dx.doi.org/10.1088/1742-5468/adb7d3}{{\em J. Stat. Mech.} (2025) 033102}.

\bibitem{QA}
V.~Alba and P.~Calabrese, {\slshape Quench action and r\'enyi entropies in integrable systems,} \href{https://link.aps.org/doi/10.1103/PhysRevB.96.115421}{{\em Phys. Rev. B} {\bfseries 96} (2017) 115421}.

\bibitem{klobas2021}
K.~Klobas and B.~Bertini, {\slshape {Entanglement dynamics in Rule 54: Exact results and quasiparticle picture},} \href{https://scipost.org/10.21468/SciPostPhys.11.6.107}{{\em SciPost Phys.} {\bfseries 11} (2021) 107}.

\bibitem{PRX}
B.~Bertini, K.~Klobas, V.~Alba, G.~Lagnese, and P.~Calabrese, {\slshape Growth of r\'enyi entropies in interacting integrable models and the breakdown of the quasiparticle picture,} \href{https://link.aps.org/doi/10.1103/PhysRevX.12.031016}{{\em Phys. Rev. X} {\bfseries 12} (2022) 031016}.

\bibitem{PhysRevLett.131.140401}
B.~Bertini, P.~Calabrese, M.~Collura, K.~Klobas, and C.~Rylands, {\slshape Nonequilibrium full counting statistics and symmetry-resolved entanglement from space-time duality,} \href{https://link.aps.org/doi/10.1103/PhysRevLett.131.140401}{{\em Phys. Rev. Lett.} {\bfseries 131} (2023) 140401}.

\bibitem{bertini2024dynamics}
B.~Bertini, K.~Klobas, M.~Collura, P.~Calabrese, and C.~Rylands, {\slshape Dynamics of charge fluctuations from asymmetric initial states,} \href{https://link.aps.org/doi/10.1103/PhysRevB.109.184312}{{\em Phys. Rev. B} {\bfseries 109} (2024) 184312}.

\bibitem{PhysRevA.98.032302}
E.~Cornfeld, M.~Goldstein, and E.~Sela, {\slshape Imbalance entanglement: Symmetry decomposition of negativity,} \href{https://link.aps.org/doi/10.1103/PhysRevA.98.032302}{{\em Phys. Rev. A} {\bfseries 98} (2018) 032302}.

\bibitem{Murciano_2021}
S.~Murciano, R.~Bonsignori, and P.~Calabrese, {\slshape Symmetry decomposition of negativity of massless free fermions,} \href{http://dx.doi.org/10.21468/SciPostPhys.10.5.111}{{\em SciPost Phys.} {\bfseries 10} (2021) }.

\bibitem{Fioretto_Mussardo}
D.~Fioretto and G.~Mussardo, {\slshape {Quantum Quenches in Integrable Field Theories},} \href{http://dx.doi.org/10.1088/1367-2630/12/5/055015}{{\em New J. Phys.} {\bfseries 12} (2010) 055015}.

\bibitem{ares2023lack}
F.~Ares, S.~Murciano, E.~Vernier, and P.~Calabrese, {\slshape {Lack of symmetry restoration after a quantum quench: An entanglement asymmetry study},} \href{https://scipost.org/10.21468/SciPostPhys.15.3.089}{{\em SciPost Phys.} {\bfseries 15} (2023) 089}.

\bibitem{ares2023}
F.~Ares, S.~Murciano, and P.~Calabrese, {\slshape {Entanglement asymmetry as a probe of symmetry breaking},} \href{https://www.nature.com/articles/s41467-023-37747-8}{{\em Nature Comm,} {\bfseries 14} (2023) 2036}.

\bibitem{Fagotti2014}
M.~Fagotti, M.~Collura, F.~H.~L. Essler, and P.~Calabrese, {\slshape Relaxation after quantum quenches in the spin-$\frac{1}{2}$ heisenberg xxz chain,} \href{https://link.aps.org/doi/10.1103/PhysRevB.89.125101}{{\em Phys. Rev. B} {\bfseries 89} (2014) 125101}.

\bibitem{Bertini_2018}
B.~Bertini, M.~Fagotti, L.~Piroli, and P.~Calabrese, {\slshape Entanglement evolution and generalised hydrodynamics: noninteracting systems,} \href{https://dx.doi.org/10.1088/1751-8121/aad82e}{{\em J. Phys. A} {\bfseries 51} (2018) 39LT01}.

\bibitem{fagotti2012evolution}
M.~Fagotti and P.~Calabrese, {\slshape Evolution of entanglement entropy following a quantum quench: Analytic results for the $xy$ chain in a transverse magnetic field,} \href{https://link.aps.org/doi/10.1103/PhysRevA.78.010306}{{\em Phys. Rev. A} {\bfseries 78} (2008) 010306}.

\bibitem{Eisler_2015}
V.~Eisler and Z.~Zimborás, {\slshape On the partial transpose of fermionic gaussian states,} \href{https://dx.doi.org/10.1088/1367-2630/17/5/053048}{{\em New J. Phys.} {\bfseries 17} (2015) 053048}.

\bibitem{Coser_2016}
A.~Coser, E.~Tonni, and P.~Calabrese, {\slshape Towards the entanglement negativity of two disjoint intervals for a one dimensional free fermion,} \href{http://dx.doi.org/10.1088/1742-5468/2016/03/033116}{{\em J. Stat. Mech.} (2016) 033116}.

\bibitem{Eisler_2007}
V.~Eisler and I.~Peschel, {\slshape Evolution of entanglement after a local quench,} \href{https://dx.doi.org/10.1088/1742-5468/2007/06/P06005}{{\em J. Stat. Mech.} (2007) P06005}.

\bibitem{Peschel_2009}
I.~Peschel and V.~Eisler, {\slshape Reduced density matrices and entanglement entropy in free lattice models,} \href{https://dx.doi.org/10.1088/1751-8113/42/50/504003}{{\em J. Phys. A} {\bfseries 42} (2009) 504003}.

\bibitem{PhysRevB.69.075111}
S.-A. Cheong and C.~L. Henley, {\slshape Many-body density matrices for free fermions,} \href{https://link.aps.org/doi/10.1103/PhysRevB.69.075111}{{\em Phys. Rev. B} {\bfseries 69} (2004) 075111}.

\bibitem{alba_carollo2021}
V.~Alba and F.~Carollo, {\slshape Spreading of correlations in markovian open quantum systems,} \href{https://link.aps.org/doi/10.1103/PhysRevB.103.L020302}{{\em Phys. Rev. B} {\bfseries 103} (2021) L020302}.

\bibitem{carollo_alba2022}
F.~Carollo and V.~Alba, {\slshape Dissipative quasiparticle picture for quadratic markovian open quantum systems,} \href{https://link.aps.org/doi/10.1103/PhysRevB.105.144305}{{\em Phys. Rev. B} {\bfseries 105} (2022) 144305}.

\bibitem{Alba_2022}
V.~Alba and F.~Carollo, {\slshape Hydrodynamics of quantum entropies in ising chains with linear dissipation,} \href{https://dx.doi.org/10.1088/1751-8121/ac48ec}{{\em J. Phys. A Math. Theor.} {\bfseries 55} (2022) 074002}.

\bibitem{carollo2022}
F.~Carollo and V.~Alba, {\slshape Entangled multiplets and spreading of quantum correlations in a continuously monitored tight-binding chain,} \href{https://link.aps.org/doi/10.1103/PhysRevB.106.L220304}{{\em Phys. Rev. B} {\bfseries 106} (2022) L220304}.

\end{thebibliography}\endgroup
\end{document}